\documentclass[journal,twoside]{IEEEtran}
\usepackage{cancel}
\usepackage{tabularx,booktabs}
\usepackage{lipsum}
\newcolumntype{C}{>{\centering\arraybackslash}X} 

\usepackage{cite}

\usepackage{wrapfig}
\ifCLASSINFOpdf
   \usepackage[pdftex]{graphicx}
   \graphicspath{{../pdf/}{../jpeg/}}
   \DeclareGraphicsExtensions{.pdf,.jpeg,.png}
\else
   \usepackage[dvips]{graphicx}
   \graphicspath{{../eps/}}
   \DeclareGraphicsExtensions{.eps}
\fi
\usepackage[cmex10]{amsmath}
\usepackage{textcomp,gensymb}
\usepackage{relsize}
\usepackage{color,soul}

\usepackage{scalerel} 
\usepackage{tikz} 
\usetikzlibrary{svg.path}

\definecolor{orcidlogocol}{HTML}{A6CE39}
\tikzset{
	orcidlogo/.pic={
		\fill[orcidlogocol] svg{M256,128c0,70.7-57.3,128-128,128C57.3,256,0,198.7,0,128C0,57.3,57.3,0,128,0C198.7,0,256,57.3,256,128z};
		\fill[white] svg{M86.3,186.2H70.9V79.1h15.4v48.4V186.2z}
		svg{M108.9,79.1h41.6c39.6,0,57,28.3,57,53.6c0,27.5-21.5,53.6-56.8,53.6h-41.8V79.1z M124.3,172.4h24.5c34.9,0,42.9-26.5,42.9-39.7c0-21.5-13.7-39.7-43.7-39.7h-23.7V172.4z}
		svg{M88.7,56.8c0,5.5-4.5,10.1-10.1,10.1c-5.6,0-10.1-4.6-10.1-10.1c0-5.6,4.5-10.1,10.1-10.1C84.2,46.7,88.7,51.3,88.7,56.8z};
	}
}

\newcommand{\orcidicon}[1]{\href{https://orcid.org/#1}{\mbox{\scalerel*{
				\begin{tikzpicture}[yscale=-1,transform shape]
				\pic{orcidlogo};
				\end{tikzpicture}
			}{|}}}}

\usepackage [hidelinks] {hyperref} 

\usepackage{algorithmic}
\usepackage[ruled,vlined]{algorithm2e} 
\usepackage{amsmath,amssymb,amsfonts}

\begin{document}
\title{5G Multi-BS Positioning:\\ A Decentralized Fusion Scheme}
\author{Sharief~Saleh\textsuperscript{\orcidicon{0000-0003-1365-417X}}\,,~\IEEEmembership{Graduate Student                     Member,~IEEE,}
        Qamar~Bader\textsuperscript{\orcidicon{0000-0002-4667-1710}}\,,~\IEEEmembership{Graduate Student Member,~IEEE}
        Mohamed~Elhabiby\textsuperscript{\orcidicon{0000-0002-1909-7506}}\,,~\IEEEmembership{Member,~IEEE}
        Aboelmagd~Noureldin\textsuperscript{\orcidicon{0000-0001-6614-7783}}\,,~\IEEEmembership{Senior Member,~IEEE}


\thanks{Sharief Saleh, Qamar Bader, and Aboelmagd Noureldin are with the Department of Electrical and Computer Engineering, Queen's University, Kingston, ON K7L 3N6, Canada, and also with the Navigation and Instrumentation (NavINST) Lab, Department of Electrical and Computer Engineering, Royal Military Collage of Canada, Kingston, ON  K7K 7B4, Canada (e-mail: sharief.saleh@queensu.ca; qamar.bader@queensu.ca; nourelda@queensu.ca).}
\thanks{Mohamed Elhabiby is with Micro Engineering Tech. Inc., and also with the Public Works Department, Ain Shams University, Cairo 11566, Egypt (e-mail: mmelhabi@ucalgary.ca).}}


\markboth{}%
{Saleh \MakeLowercase{\textit{et al.}}: 5G Multi-BS Positioning: A Decentralized Fusion Scheme}





\maketitle
\bstctlcite{IEEEexample:BSTcontrol} 

\begin{abstract}
Fifth-generation (5G) networks are expected to provide high-precision positioning estimation utilizing mmWave signals in urban and downtown areas. In such areas, 5G base stations (BSs) will be densely deployed, allowing for line-of-sight (LOS) communications between the user equipment (UE) and multiple BSs at the same time. Having access to a plethora of measurement sources grants the need for optimal integration between the BSs to have an accurate and precise positioning solution. Traditionally, 5G multi-BS fusion is conducted via an extended Kalman filter (EKF), that directly utilizes range and angle measurements in a centralized integration scheme. Such measurements have a non-linear relationship with the positioning states of the filter, giving rise to linearization errors. Counter to the common belief, an unscented Kalman filter (UKF) will fail to totally eradicate such linearization errors. In this paper, we argue that a decentralized integration between 5G BSs would fully avoid linearization errors within the filter and significantly enhance the positioning performance. This is done by fusing position measurements instead of directly fusing range and angle measurements, which inherently leads to a linear measurement model by design. The proposed decentralized KF method was evaluated in a quasi-real simulation setup provided by Siradel using a real trajectory in Downtown Toronto. The experiments compared the performance of decentralized KF integration to that of centralized EKF and UKF integration schemes. It was shown that the proposed method outperformed both UKF and EKF implementations in multiple scenarios as it significantly decreased the RMS and maximum 2D positioning errors, achieving decimeter-level accuracy for $93.9\%$ of the time.
\end{abstract}

\begin{IEEEkeywords}
5G; angle of departure (AOD); autonomous vehicles (AVs); decentralized sensor fusion; Kalman filter (KF); localization; mm-Wave; positioning; round trip time (RTT).
\end{IEEEkeywords}

\section{Introduction}
\IEEEPARstart{P}{recise} positioning and navigation solutions are becoming of considerable importance in recent years due to the growing research in highly-intelligent internet of things (IoT) applications, context-aware services, and location-based services (LBS). Additionally, precise position information is necessary for safety-critical applications including self-driving cars; which require higher levels of autonomy. Such systems compel centimeter-level of absolute and relative positioning accuracy, low latency, and high reliability\cite{Reid_2019}. Wireless-based positioning has been utilized for decades to provide positioning solutions for applications where high accuracy is subordinate. Such technologies utilize time-, angle-, and power-based measurements to find the user equipment's (UE) position by means of trilateration, triangulation, and hybrid positioning techniques \cite{5GPos}. Trilateration employs ranges acquired by three or more base stations (BSs) to obtain the position of the UE. Triangulation, on the other hand, exploits multiple relative angle measurements to estimate the position of the UE. Lastly, hybrid positioning approaches utilize both range and angle measurements to get the location of the user, which can be achieved through a single BS. Global navigation satellite systems (GNSS) for instance, can provide a few-meter level of accuracy in standalone operation and down to cm-level of accuracy when aided by real-time kinematic (RTK) technology \cite{RTK}. However, due to multi-path and signal blockage, GNSS-based positioning solutions deteriorate in urban canyon environments. On the other hand, WiFi and ultra-wideband (UWB) are capable of attaining sub-meter level of accuracy, yet, both technologies suffer from short-range operation ranging between $50$ m to $200$ m\cite{UWB} \cite{WiFi}, rendering them advantageous only in indoor environments. On the contrary, 5G new radio (NR) small-cells are expected to be densely deployed in deep-urban for communication purposes with an inter-cell distance of $200$ to $500$ meters \cite{merits}. This will increase the chance of line-of-sight (LoS) communications with the user, which is important for LOS-based positioning algorithms. Furthermore, 5G mmWave positioning signals will span a bandwidth of up to $400$ MHz, yielding accurate time-based measurements estimation, such as time of arrival (TOA), round trip time (RTT), and time difference of arrival (TDOA), as well as high multi-path (MP) resolvability in the time domain. Additionally, the 5G frequency range 2 (FR2) operates on high-frequency mmWave signals spanning from $28$ GHz to $52$ GHz, allowing for the emergence of massive MIMO. Thus, 5G receivers will comprise a large number of antennas enabling accurate angle-based measurements such as angle of arrival (AOA) and angle of departure (AOD), as well as high MP resolvability in the angle domain.

\IEEEpubidadjcol  
In order to optimize the estimation of the UE's position considering noisy range and angle measurements, optimal state estimators are needed. The most commonly used filters are the Kalman filter (KF) \cite{KF}, the extended Kalman filter (EKF) \cite{EKF} and the unscented Kalman filter (UKF) \cite{ukf}. While KF assumes the linearity of both the state transition and the observation models, the remaining stated filters do not oblige the same requirement. Additionally, KF, EKF, and UKF assume normally distributed process noise and measurement noise to function optimally. Breaching any of these requirements will revoke the filters' optimality. The aforementioned filters can be utilized in centralized or decentralized state estimation schemes, also known as tightly coupled (TC) and loosely coupled (LC) integration schemes, respectively. TC and LC integration schemes are well-known in the GNSS/INS community. Centralized/TC schemes integrate all technologies/BSs on the measurement level within a single-stage filter. Decentralized/LC schemes, on the other hand, involve multi-stage positioning where each technology/BS computes its own estimate of the position, and then integration on the position level takes place in a cascaded manner. As will be seen in Section \ref{LR}, all methods proposed in the literature utilize a centralized integration scheme. Such a scheme inherently revokes the optimality of the filters, in the case of 5G positioning, as the relationship between the states and the measurements is non-linear. Simulations of the detrimental effects of centralized integration will be shown in Section \ref{centralized}. It is then shown that the decentralized integration of multi-BSs totally avoids the linearization issues as it adapts a linear measurement model by design. Hence, the contributions of this paper are as follows:
\begin{enumerate}
  \item We present an extensive analysis of the drawbacks of centralized/TC integration schemes.
  \item A novel decentralized/LC multi-BS 5G positioning approach is proposed, which will avoid the disadvantages of centralized integration schemes.
  \item For validation, we assess three Bayesian estimators, namely, LC-KF, TC-EKF, and TC-UKF in a quasi-real testing environment.
\end{enumerate}
The remainder of the paper is organized as follows: Section II presents related works on 5G positioning using optimal estimators. Section III establishes the system model and  Section IV exhibits the various implementations of Kalman filters. Section V thoroughly analyzes the centralized fusion scheme errors and proposes 5G decentralized integration as a strong contender. Section VI provides details on the experimental setup and presents the results along with a discussion. Finally, Section VII concludes the paper.

\section{Literature Review} \label{LR}
Various works have investigated LOS 5G positioning using different estimation techniques. Yet, a common theme that is prevalent throughout the literature is the use of centralized integration schemes. It is worth noting that methods that rely on trilateration or triangulation are considered centralized by nature. Authors in \cite{TrainTOAOA} used EKF to estimate the UE's position employing AOD and TOA measurements, giving rise to high linearization errors due to the use of highly non-linear observation models, achieving sub-meter accuracy for $75\%$ of the time. In \cite{TrainTDOA1,TrainTDOA2} an accuracy of the sub-meter level was accomplished for $90\%$ of the time through utilizing TDOA trilateration. However, this was achieved assuming that the UE is connected to three BSs simultaneously at all times, which is an impractical assumption. Authors in \cite{TrainTDOAAOA} proposed a hybrid positioning scheme based on TDOA, and AOA using an EKF, achieving a sub-3-meter level of accuracy for $95\%$ of the time. Nevertheless, it still endures the same limitations as observed in previous works. In \cite{TrainTDOAAOA}, authors have integrated TDOA and AOA measurements in a centralized manner and achieved sub-3 meters of accuracy $95\%$ of the time. The work in \cite{TrainNonLinear} has mitigated some of these limitations by endorsing a non-linear state transition model (NLSTM) as an alternative to the constant velocity model. This model makes use of the forward velocity of the vehicle and its heading angle to estimate the vehicle's position. The proposed NLSTM surpassed former works, attaining sub-2.3 meters of accuracy for $95\%$ of the time. Inspired by \cite{TrainNonLinear}, authors in \cite{trivedi_localization_2021} utilized the same NLSTM model but with centralized integration of TOA, AOA, and heading measurements instead. The heading of the vehicle measurement was artificially computed by combining AOA, AOD, and TOA measurements. The use of TOA measurements without proper estimation of the clock bias between the UE and the BS is not practical. Additionally, the proposed method utilizes correlated measurements without proper manipulation of the measurement covariance matrix, which leads to the sub-optimality of the EKF. The proposed method was able to achieve $0.34$ m error for $95\%$ of the time by using two BSs only. Authors in \cite{5GHybrid2} integrate rough TOA estimates in a centralized fashion to directly position the UE via a compressed sensing approach. The proposed method achieved a sub-meter level of accuracy for around $90\%$ of the time. An EKF-based triangulation scheme based on three AOD measurements was proposed by \cite{rastorgueva-foi_beam-based_2018}. The method also estimates the orientation bias of the BS due to errors in installation. The proposed triangulation method was able to achieve a sub-meter level of accuracy for $88\%$ of the time. Yet, the assumption of having access to three BSs that are in LOS with the UE is usually impractical. The work reported in \cite{gertzell_5g_2020} proposes a least-squares approach to directly integrate TDOA and AOD measurements from multiple BSs. The minimum number of LOS BSs needed for such an approach is two BSs, which is not usually available to the UE. Moreover, LS approaches neither have the optimal weighting capabilities nor the tracking  capability of KF implementations. The proposed LS method was reported to have the ability to estimate the position with a sub-meter level of accuracy without reporting the confidence interval on such an assumption. Authors in \cite{ko_high-speed_2022} propose two centralized EKF-based trilateration solutions that utilize TDOA measurements. The first method proposes a modified implementation of the EKF, where they added additional noise to the measurement's covariance matrix and the state covariance matrix to account for the non-linearities and the non-gaussian distribution of the noises. The second method proposes a deep-EKF scheme, which utilizes neural networks (NN) to replace the innovation sequence of the modified EKF to better capture non-linearities. The results of the proposed method show a clear improvement in the RMS error compared to conventional EKF implementations, yet, it did not fully report the statistics of the solution. The work described in \cite{xhafa_evaluation_2022} directly fuses TDOA and AOA measurements to estimate the position of the UE via a Gauss-Newton LS method. The proposed method also includes a BS exclusion mechanism based on NLOS detection and synchronization anomalies. The proposed method shows poor positioning performance due to the use of cm-waves. Limited work was done on the use of UKF for 5G standalone positioning. In \cite{5GSync2} a UKF-based estimator was employed to simultaneously estimate AOA, TOA, and the position of the UE through a centralized scheme, accomplishing a sub-meter level of accuracy. Nevertheless, they have assumed BSs are deployed every $50$ m, which does not follow the standard for small cells deployed in urban areas as seen in \cite{whitepaper}. It is worth noting that other works proposed joint positioning and synchronization schemes like \cite{5GSync1,bai_toa-aoa_2021}, and others have proposed joint estimation of UE position and positioning measurables as in \cite{rastorgueva-foi_localization_2018,shahmansoori_tracking_2019}. Moreover, works in \cite{5GSync2,5GSync3} have proposed joint estimation of UE position, positioning measurables as well as UE-BS synchronization. Works that jointly estimate positioning measurables usually utilize a multi-stage filtering approach, where positioning measurables are first estimated through an EKF/UKF filter, then positioning takes place in a separate filter. Such schemes are not to be considered decentralized schemes, as the positioning solution is still computed within a single centralized filter. Works related to UE-BS synchronization conduct time bias estimation concurrently with UE position and orientation estimation, and are also conducted in a centralized fashion. In conclusion, all of the aforementioned works utilize centralized integration schemes to integrate the BSs' various raw time and angle measurements. This will prove detrimental to the accuracy of the 5G positioning solution. 

\section{System Model}
5G NR has enhanced the LTE's downlink (DL) positioning reference signal (PRS) and uplink (UL) sounding reference signal (SRS) to allow for more precise positioning measurements. With the aid of these signals, TOA, RTT, and TDOA are computed through correlation with the configured reference signal. Moreover, AOD is inherently computed within the beamforming training sequence by comparing the received power of multiple PRS signals. In this paper, the authors consider positioning at the UE side, and thus, DL-PRS and UL-SRS signals are utilized to compute DL-AOD and RTT measurables. Such measurables are then used to compute the relative distance between the UE and the BS $(r)$, as well as the azimuth angle $(\theta)$ as seen in (\ref{Range}) and (\ref{Angle}).

\begin{equation}\label{Range}
    \begin{split}
         r&=\sqrt{\Delta x^2+ \Delta y^2 + \Delta z^2}\\
    \end{split}
\end{equation}

\begin{equation}\label{Angle}
    \theta=\arctan\left(\frac{\Delta y}{\Delta x}\right)
\end{equation}

\begin{figure}[h!]
\centering
\includegraphics[trim=0pt 0pt 0pt 0pt,clip,width=\columnwidth]{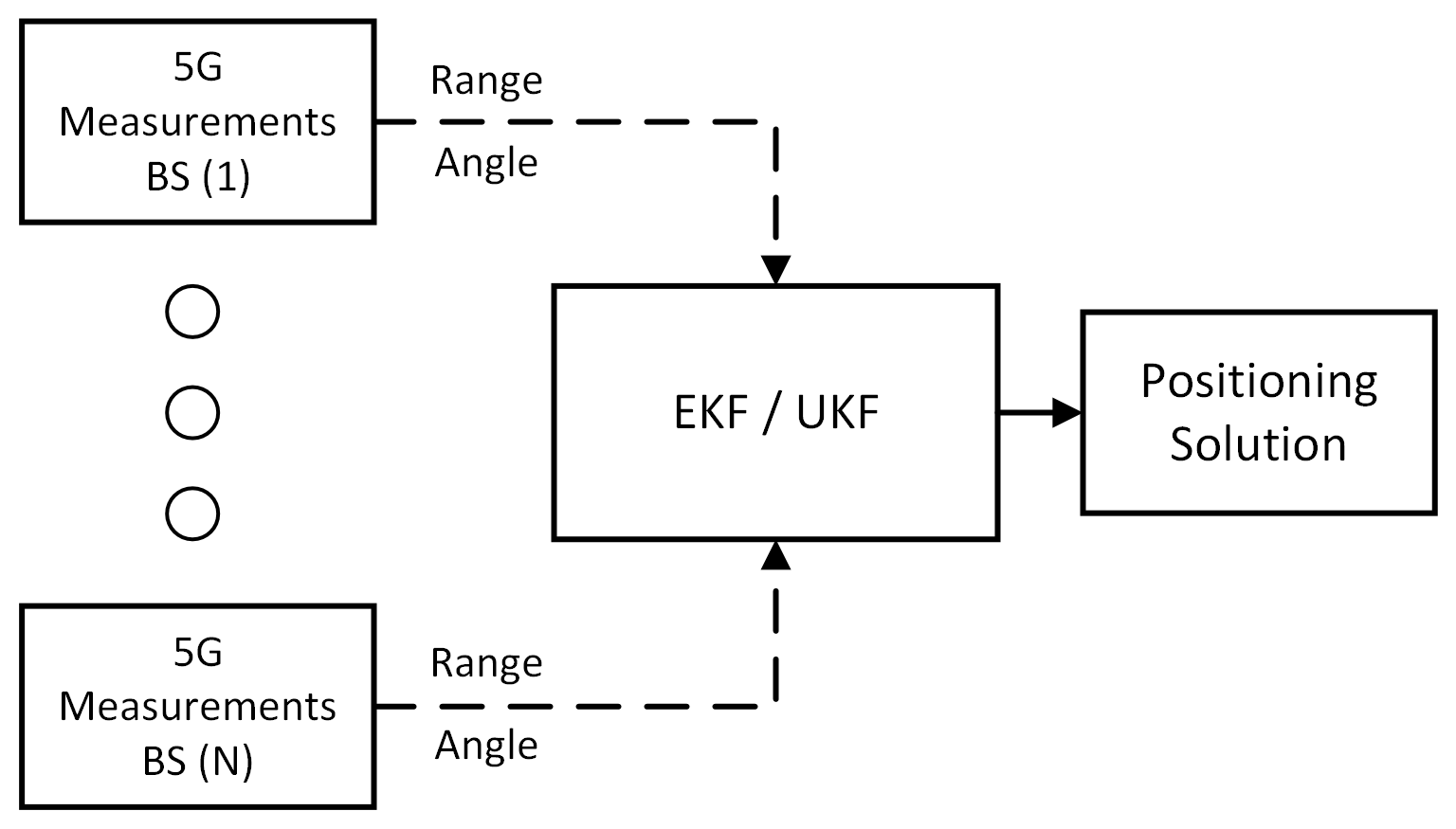}
\caption{Block diagram of centralized 5G multi-BS integration.}
\label{TC}
\end{figure}

\begin{figure}[h!]
\centering
\includegraphics[trim=0pt 0pt 0pt 0pt,clip,width=\columnwidth]{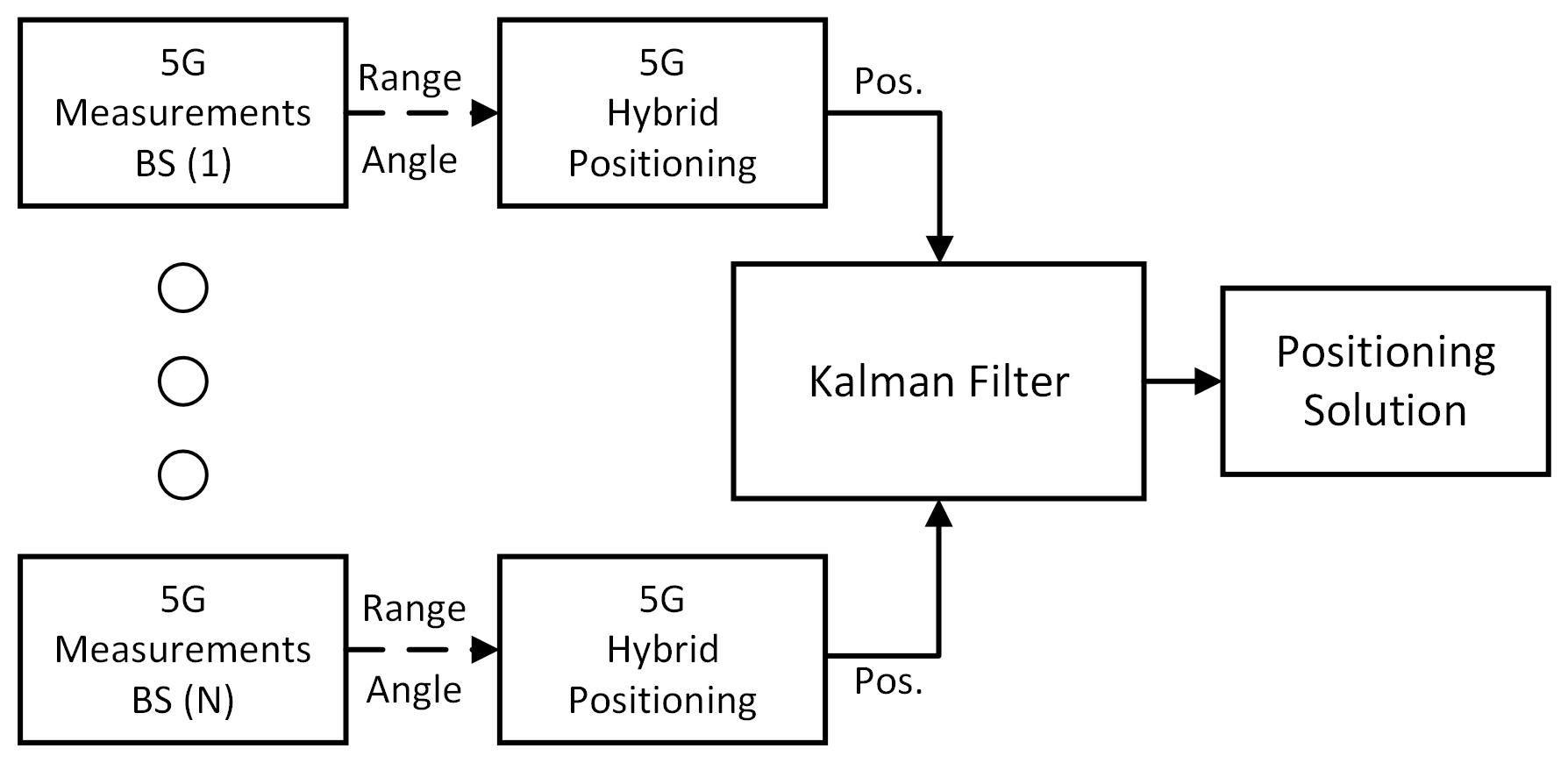}
\caption{Block diagram of decentralized 5G multi-BS integration.}
\label{LC}
\end{figure}

\noindent  Where $(\Delta x, \Delta y, \Delta z)$ denote the difference between the UE's position $(x,y,z)$ and the BS position $(x_{BS},y_{BS},z_{BS})$. Next, the range and angle measurements offered by each connected BS can either be fused in a centralized or decentralized fashion. In centralized fusion, all ranges and angles are fused in a single filter to compute the UE's position as seen in Fig.~\ref{TC}. On the other hand, in decentralized fusion, the UE first computes multiple position estimates by using individual BS measurables through an open-loop geometry-based calculation. Subsequently, an optimal estimator fuses these position estimates into a single position estimate as seen in Fig.~\ref{LC}.
\noindent Finally, to explore the true potential of 5G-NR's small-cell densification, the setting of autonomous driving was set to be in downtown areas where urban canyons and high-rise buildings surround the vehicle. The simulation setup, shown in Section \ref{Exp Setup}, was designed as per 3GPP release 16 standards for deep-urban environments, which entails that BSs will be deployed $200-500$ m apart \cite{Course}. Designing such a setting showed that a UE might be connected to a maximum of three BSs at a time with a probability of $30\%$. It also showed that the UE will have access to two BSs and sometimes a single BS for $46\%$ and $21\%$ of the time, respectively. Finally, a total loss of LOS connection to all BSs occurs in rare cases with a probability of $3\%$ of the time. Thus, an NLOS detection algorithm based on range comparison between RTT and RSS measurements, proposed in \cite{NLOS}, will be utilized to exclude NLOS BSs. In a nutshell, the method exploits discrepancies between time-based and power-based measurements to detect NLOS communications. This is done by comparing the difference in range computed by TOA and RSS measurements to a preset threshold as seen in (\ref{NLOS}).
\begin{equation}\label{NLOS}
 \text{LOS}=
\begin{cases}
\text{True}, & r_{TOA}-r_{RSS} \leq \varepsilon\\
\text{False}, &\text{otherwise.}
\end{cases}
\end{equation}
\noindent Where $r_{TOA}$ and $r_{RSS}$ are the time-based and power-based range measurements, respectively, and $\varepsilon$ is the preset threshold set to $80$ m, as per \cite{NLOS}.
\vspace{-3pt}

\section{Kalman Filter Realizations}
The Kalman filter is an optimal iterative estimator for linear systems with Gaussian noise, illustrated in Alg. \ref{KF Alg}. With these two assumptions, the KF is able to propagate, transform, and manipulate the states' probability density function (PDF) by only tracking their mean and covariance expressed by $\boldsymbol{x}$ and $\boldsymbol{P}$, respectively. The KF consists of two consecutive stages of state prediction and correction. In the state prediction stage, the KF propagates the posterior state estimate $\boldsymbol{x}_{k-1}^+$ and the covariance matrix $\boldsymbol{P}_{k-1}^+$ from the previous epoch, $k-1$, to the current epoch, $k$, via the transition matrix $\boldsymbol{\Phi}=\boldsymbol{I}+\boldsymbol{F}$. Where $\boldsymbol{I}$ is an identity matrix and $\boldsymbol{F}$ is the transition model. The outputs of such transformation are the a priori state estimate $\boldsymbol{x}_{k}^-$ and the covariance matrix $\boldsymbol{P}_{k}^-$. Noise caused by external sources within the transition model is described by the process covariance matrix $\boldsymbol{Q}$. The process noise is added to the a priori covariance matrix with the aid of the noise coupling matrix $\boldsymbol{G}$. The state correction stage utilizes the measurement vector $\boldsymbol{z}_k$ along with the Kalman gain $\boldsymbol{K}_k$ to update the a priori state and covariance estimate into a posteriori estimates $\boldsymbol{x}^+_{k}$ and $\boldsymbol{P}^+_{k}$. The Kalman gain $\boldsymbol{K}_k$ is computed with the aid of $\boldsymbol{P}_{k}^-$ and the measurement covariance matrix $\boldsymbol{R}$. Moreover, the measurement observation matrix $\boldsymbol{H}_k$ is used to transform $\boldsymbol{P}_{k}^-$ and $\boldsymbol{x}^-_{k}$ from the state domain to the measurement domain and vice versa.

It is worth noting that the main difference between linear KF and EKF is that EKF deals with non-linear transition models and non-linear state-measurement models. In case of encountering non-linearity in the transition model, $f(\boldsymbol{x}^+_{k-1})$ is used to translate the state instead of $\boldsymbol{\Phi} \boldsymbol{x}^+_{k-1}$, and $\boldsymbol{F}$ is computed by finding the Jacobian of $f(\boldsymbol{x})$. Likewise, in case of encountering non-linearity between the states and the measurements, then $h(\boldsymbol{x}^-_k)$ is used instead of $\boldsymbol{H}_k \boldsymbol{x}^-_{k}$ to transform the states to measurements, and $\boldsymbol{H}_k$ is computed by finding the Jacobian of $h(\boldsymbol{x})$. It is worth noting that $\boldsymbol{\Phi}$ and $\boldsymbol{H}_k$ are still utilized to transition and transform the state covariance matrix $\boldsymbol{P}$, respectively.

The use of UKF resolves the non-linear approximation drawbacks of the EKF by transforming the state's PDFs through non-linear functions. The UKF conducts PDF transformation by generating $2N+1$ sigma points $\boldsymbol{\chi}_k$, where $N$ is the number of connected BSs. Then, the UKF propagates the individual  $\boldsymbol{\chi}_k$ points through the non-linear function and finally computes the mean and covariance of the propagated points to find $\boldsymbol{x}$ and $\boldsymbol{P}$, respectively. The UKF algorithm is illustrated in Alg. \ref{UKF Alg}.

\begin{algorithm}[t!]
\SetAlgoLined
 initialization: $\boldsymbol{x}^+_{0},\boldsymbol{P}^+_{0},\boldsymbol{Q},\boldsymbol{R},k=0$\\
 \While{positioning}{
  $k=k+1$\\ \vspace{0.5pt}
  $\boldsymbol{x}^-_{k}=\boldsymbol{\Phi} \boldsymbol{x}^+_{k-1}$\\ \vspace{0.5pt}
  $\boldsymbol{P}^-_{k}=\boldsymbol{\Phi} \boldsymbol{P}^+_{k-1}\boldsymbol{\Phi}^T + \boldsymbol{GQG}^T$\\ \vspace{0.5pt}
  $\boldsymbol{K}_{k}=\boldsymbol{P}^-_{k}\boldsymbol{H}^T_k(\boldsymbol{H}_k \boldsymbol{P}^-_{k} \boldsymbol{H}^T_k +\boldsymbol{R})$\\ \vspace{0.5pt}
  $\boldsymbol{x}^+_{k}=\boldsymbol{x}^-_{k}+\boldsymbol{K}_k(\boldsymbol{z}_k-\boldsymbol{H}_k\boldsymbol{x}^-_{k})$\\ \vspace{0.5pt}
  $\boldsymbol{P}^+_{k}=\boldsymbol{P}^-_{k}-\boldsymbol{K}_k\boldsymbol{H}_k\boldsymbol{P}^-_{k}$\\ \vspace{0.5pt}}
 \caption{Kalman Filter}
 \label{KF Alg}
\end{algorithm}

\begin{algorithm}[t]
\SetAlgoLined
 initialization: $\boldsymbol{x}^+_{0},\boldsymbol{P}^+_{0},\boldsymbol{Q},\boldsymbol{R}, W_m^i, \lambda, k=0$\\
 \While{positioning}{
  $k=k+1$\\ \vspace{1pt}
  $\boldsymbol\chi_k^{-(i)} = \left[\hat{\boldsymbol{x}}_{k-1}^+ \pm \sqrt{\lambda + N} \left[ \sqrt{\boldsymbol P_{k-1}^+}  \right] \right] $\\\vspace{1pt}
  
  $\boldsymbol{x}_k^{(i)} = f\left(\boldsymbol\chi_k^{-(i)}\right); i = 0, 1, \dots, 2N$\\\vspace{1pt}
  
  $\hat{\boldsymbol{x}}_k^{-} = \sum\limits_{i=0}^{2N} W_m^i \boldsymbol{x}_k^{(i)}$\\\vspace{1 pt}
  
  $\boldsymbol{P}_k^{-} = \sum\limits_{i=0}^{2N} W_m^i \left(\boldsymbol{x}_k^{(i)} - \hat{\boldsymbol{x}}_k^{-}\right)\left(\boldsymbol{x}_k^{(i)} - \hat{\boldsymbol{x}}_k^{-}\right)^T \hspace{-5pt}+ \boldsymbol{Q} $\\\vspace{1pt}
  
  $\boldsymbol{z}_k^{(i)} = h\left(\boldsymbol{x}_k^{(i)}\right)$\\\vspace{1pt}
  
  $\hat{\boldsymbol{z}}_k = \sum\limits_{i=0}^{2N} W_m^i \boldsymbol{z}_k^{(i)}$\\\vspace{1pt}
  
  $\boldsymbol{P}_z = \sum\limits_{i=0}^{2N} W_m^i \left(\boldsymbol{z}_k^{(i)} - \hat{\boldsymbol{z}}_k^{-}\right)\left(\boldsymbol{z}_k^{(i)} - \hat{\boldsymbol{z}}_k^{-}\right)^T \hspace{-5pt} + \boldsymbol{R}$\\\vspace{1pt}
  
  $\boldsymbol{P}_{xz} = \sum\limits_{i=0}^{2N} W_m^i \left(\boldsymbol{x}_k^{(i)} - \hat{\boldsymbol{x}}_k^{-}\right)\left(\boldsymbol{z}_k^{(i)} - \hat{\boldsymbol{z}}_k^{-}\right)^T$\\\vspace{1pt}
  
  $\boldsymbol {K}_k = \boldsymbol {P}_{xz}\boldsymbol {P}_z^{-1}$\\\vspace{1pt}

  $\hat{\boldsymbol{x}}^+_{k}=\boldsymbol{x}^-_{k}+\boldsymbol{K}_k\left(\boldsymbol{z}_k+\hat{\boldsymbol {z}}_k\right)$\\ \vspace{1pt}
  
  $\boldsymbol{P}^+_{k}=\boldsymbol{P}^-_{k}+\boldsymbol{K}_k\boldsymbol{P}^-_{k}\boldsymbol{K}_k^T$\\ \vspace{1pt}}
 \caption{Unscented Kalman Filter}
 \label{UKF Alg}
\end{algorithm}

\section{5G Multi-BS Sensor Fusion}
In this section, both the traditional centralized integration scheme and the proposed decentralized fusion scheme will be presented. The transition model is shared between both schemes, hence, it will be introduced first. Next, the linearization errors of the centralized scheme will be displayed and thoroughly discussed. Finally, the proposed decentralized scheme will be presented.

\subsection{States and State Transition Process}
The navigation states considered in all methodologies presented in this paper comprise the estimated 2D position and velocity of the UE, i.e., $\boldsymbol{x}=[x, y, {v_x}, {v_y}]^T$. Likewise, the state transition process for all KF implementations in this work considers a constant velocity model, yielding a linear state transition model as seen in (\ref{F}).

\begin{equation} \label{F}
\boldsymbol{\Phi}=\begin{bmatrix}
1 & 0 &\Delta t & 0\\
0 & 1 & 0 &\Delta t\\
0 & 0 & 1 & 0\\
0 & 0 & 0 & 1
\end{bmatrix}
\end{equation}

The process covariance matrix $\boldsymbol{Q}$ comprises the errors arising due to violating the constant velocity model by either changing the orientation of the vehicle or by accelerating~/~decelerating. Thus, $\boldsymbol{Q}$ was set by empirically computing the average variance of the velocity of vehicles driving in a downtown setting. The experimental setup used for such empirical tuning of $\boldsymbol{Q}$ is detailed in Section VI. On the other hand, $\boldsymbol{R}$ comprises measurement errors. For centralized integration, $\boldsymbol{R}$ comprises the noise variance of the range and angle measurements, which were empirically tuned to $1$ mm and $0.1\degree$, respectively. The empirical tuning was done based on the resulting positioning RMS error from the whole trajectory. In decentralized integration, however, $\boldsymbol{R}$ is the noise variance of the position estimates from each BS. Hence, $\boldsymbol{R}$ was set to be $1$ cm for both $x$ and $y$ measurements, based on empirical tuning.

\subsection{Centralized Integration Linearization Errors} \label{centralized}
In centralized 5G multi-BS sensor fusion, the raw range and angle measurements are directly fused together in the fusion filter as seen in Fig.~\ref{TC}. The measurement vector $\boldsymbol{z}$ for both EKF and UKF centralized implementations is as follows: $\boldsymbol{z}~=~[r_1, \theta_1, \dots, r_N, \theta_N]^T$. The non-linear measurement model of the system for both EKF and UKF is expressed in (\ref{h}). Furthermore, the Jacobian matrix $\boldsymbol{H}_k \in \mathbb{R}^{2N \times 4}$ is computed using Taylor series expansion as seen in (\ref{HEKF}). It is worth noting that $\boldsymbol{H}_k$ is only utilized for centralized EKF implementation.

\begin{equation} \label{h}
h(\boldsymbol{x})=\begin{bmatrix}
\sqrt{(\Delta x_k[n])^2 + (\Delta y_k[n])^2}\\
\arctan\left(\frac{\Delta y_k[n]}{\Delta x_k[n]}\right)
\end{bmatrix}
\end{equation}

\begin{equation} \label{HEKF}
\boldsymbol{H}_k=\begin{bmatrix}
\frac{dx_1}{r_1} & \frac{dy_1}{r_1} & 0 & 0\\
\frac{-dy_1}{r_1^2} & \frac{dx_1}{r_1^2} & 0 & 0\\
\vdots & \vdots & \vdots & \vdots\\
\frac{dx_N}{r_N} & \frac{dy_N}{r_N} & 0 & 0\\
\frac{-dy_N}{r_N^2} & \frac{dx_N}{r_N^2} & 0 & 0
\end{bmatrix}
\end{equation}

Linearization errors are inevitable to occur in the centralized integration of 5G BSs using raw range and angle measurements. This is mainly caused due to the elimination of the higher-order terms during the linearization of $h(\boldsymbol{x})$. The linearized $\boldsymbol{H}$ is used in multiple places in Kalman filter implementations as seen in Alg. \ref{KF Alg}. It is therefore important to study the negative effects imposed by each use of such a linearized version of the measurement observation matrix. It is worth pointing out that a common mistake that is usually encountered in the literature while implementing EKF  is the use of $\boldsymbol{H}$ instead of $h(\boldsymbol{x})$ within the innovation sequence, i.e., the use of $\boldsymbol{z}-\boldsymbol{H}\boldsymbol{x}$ instead of $\boldsymbol{z}-h(\boldsymbol{x})$. Such a mistake will bias the rather unbiased filter, leading to extra unnecessary errors. With that being said, in EKF implementations, the observation matrix is solely used in the computation of the Kalman gain $\boldsymbol{K}$, which has two purposes in the correction process. First, it optimally weights the individual measurement errors in the innovation sequence. Second, it transforms the innovation sequence from the measurement domain to the state domain. The observation matrix $\boldsymbol{H}$ has a role in both functions. First, $\boldsymbol{H}$ is utilized to find the innovation covariance matrix $(\boldsymbol{H} \boldsymbol{P}^- \boldsymbol{H}^T +\boldsymbol{R})$. The innovation covariance matrix is essential in the weighting process of the EKF. Therefore, linearization errors in $\boldsymbol{H}$ would lead to a filter that is either overconfident or underconfident about the innovation sequence. It is worth noting that the UKF implementation is free of such errors due to using sigma points to compute the innovation covariance matrix. This will lead to a more accurate weighting of the measurements in UKF implementations. The second use of the observation matrix is to help with transforming the innovation residual errors from the measurement domain to the state domain. This is done through the use of $\boldsymbol{H}^T$. Meaning, the non-linear relationship between the states and the measurements is reduced to a linear multiplication between $\boldsymbol{H}^T$ and $\boldsymbol{z}-h(\boldsymbol{x})$. Such simplification of the non-linear relationship will induce high positioning errors in centralized EKF implementations. It is worth noting that the UKF centralized implementation will also suffer from such transformation errors, despite the usage of sigma points to compute the Kalman gain. That is due to the fact that the UKF still utilizes a linear multiplication between $\boldsymbol{K}$ and $\boldsymbol{z}-h(\boldsymbol{x})$ to perform the non-linear transformation. Thus, counter to the common belief, the UKF is not totally immune to linearization errors. As a matter of fact, the authors claim that all KF implementations and variations will suffer from linearization errors to some degree due to the inherent fundamental design of Kalman filters.

\begin{figure}[t!]
	\centering
	\includegraphics[trim=100pt 240 125pt 245pt,clip,width=1\columnwidth]{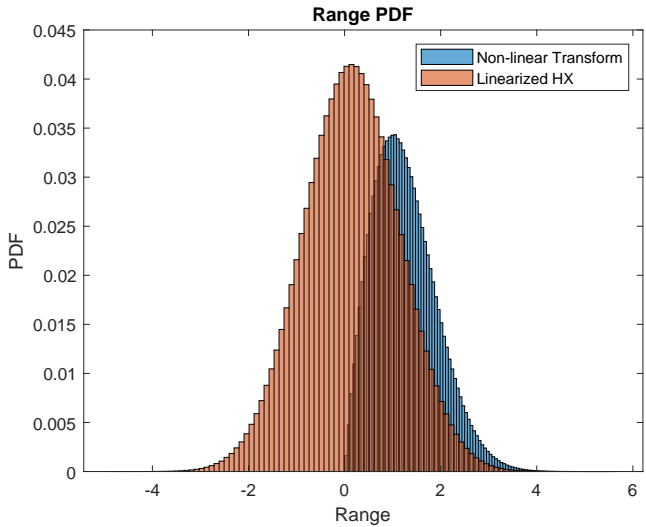}
	\DeclareGraphicsExtensions.
	\caption{Transformation of position PDF to range PDF at close BS-UE range ($\Delta x=\Delta y=0.1$ m).}
	\label{Rclose}
\end{figure}

\begin{figure}[t!]
	\centering
	\includegraphics[trim=100pt 240 125pt 245pt,clip,width=1\columnwidth]{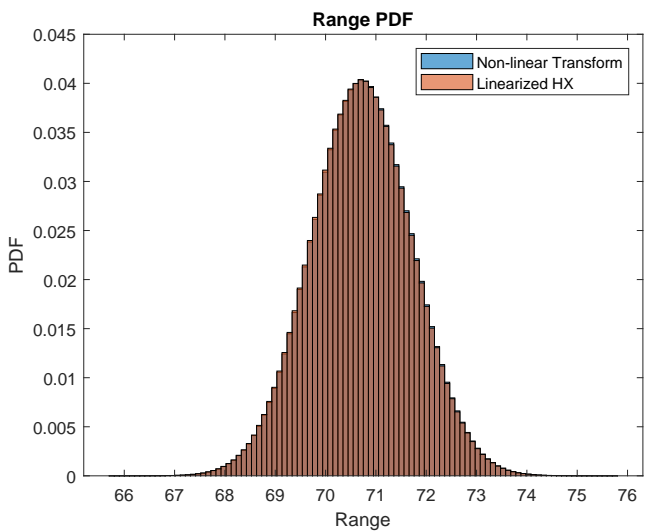}
	\DeclareGraphicsExtensions.
	\caption{Transformation of position PDF to range PDF at far BS-UE range ($\Delta x=\Delta y=50$ m).}
	\label{Rfar}
\end{figure}

To further study the discussed linearization errors, multiple theoretical simulations were conducted. The simulations address both weighting and transformation errors for both range and angle measurements. In order to simulate weighting errors, a random number generator was utilized to generate a cloud of normally distributed UE positions with a set mean and variance. The point cloud was then passed once through the non-linear measurement model $h(\boldsymbol{x})$ and through the linearized measurement model $\boldsymbol{Hx}$. Finally, the resulting transformed PDFs are displayed for comparison. The PDF resulting from the non-linear transformation acts as the ground truth for what the distribution should ideally look like. Of course, the non-linearly transformed distributions are not expected to be Gaussian as opposed to a linearly transformed distribution. Yet, it is still expected that the linearly transformed distribution would be as close as possible to the non-linearly transformed distribution to have a fair weighting of measurements. Figures~\ref{Rclose}~and~\ref{Rfar} show the simulation for range measurement's PDF transformation at close range and far range, respectively. It can be seen that at close ranges, the linearly transformed PDF does not mimic the non-linearly transformed PDF, which was expected according to \cite{DTCM}. More importantly, it can be seen that the linearly transformed PDF possesses higher variance compared to the real variance of non-linear transformation, meaning that the filter will underestimate the correction update. On the other hand, it can be seen that at far ranges, both PDFs are almost identical, resulting in the proper weighting of the measurements. 

\begin{figure}[h!]
	\centering
	\includegraphics[trim=100pt 240 125pt 245pt,clip,width=1\columnwidth]{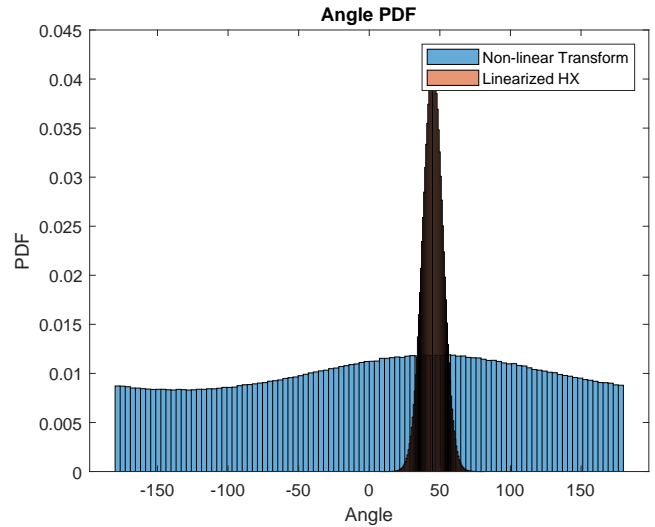}
	\DeclareGraphicsExtensions.
	\caption{Transformation of position PDF to angle PDF at close BS-UE range ($\Delta x=\Delta y=0.1$ m).}
	\label{Aclose}
\end{figure}

\begin{figure}[h!]
	\centering
	\includegraphics[trim=100pt 240 125pt 245pt,clip,width=1\columnwidth]{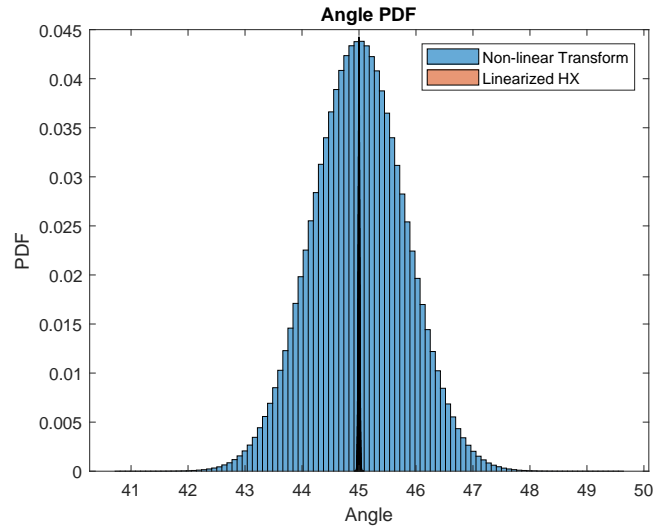}
	\DeclareGraphicsExtensions.
	\caption{Transformation of position PDF to angle PDF at far BS-UE range ($\Delta x=\Delta y=50$ m).}
	\label{Afar}
\end{figure}

Figures~\ref{Aclose}~and~\ref{Afar} present the simulation for the transformation of the PDFs of angle measurements at close range and far range, respectively. It can be seen that the linearly transformed PDF is heavily overconfident in both scenarios compared to their non-linearly transformed counterparts. Such overconfidence in measurements is dangerous in real-life navigation especially if the update is expected to be biased as well due to the transformation errors.

\begin{figure}[t!]
	\centering
	\includegraphics[width=1\columnwidth]{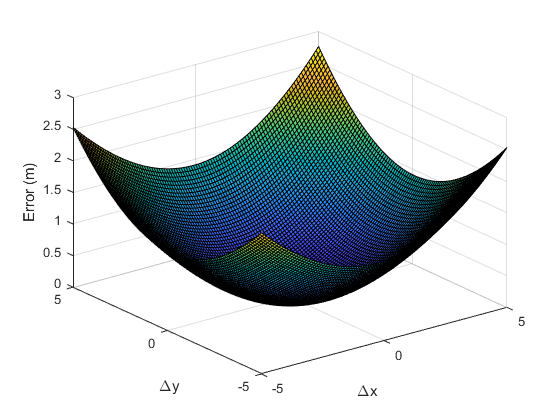}
	\DeclareGraphicsExtensions.
	\caption{Range linearization error at close BS-UE range ($\Delta x=\Delta y=1$ m).}
	\label{RcloseT}
\end{figure}

\begin{figure}[t!]
	\centering
	\includegraphics[width=1\columnwidth]{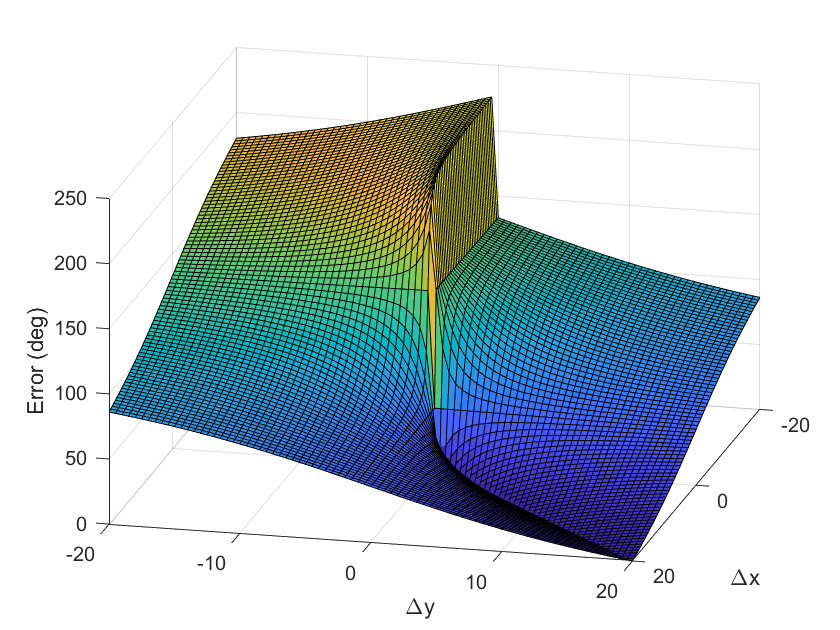}
	\DeclareGraphicsExtensions.
	\caption{Angle linearization error at $45\degree$ orientation.}
	\label{AT}
\end{figure}

In order to simulate transformation errors, a mesh of UE positions was created and utilized to compute the range and angle through the non-linear measurement model $h(\boldsymbol{x})$ and through the linearized measurement model $\boldsymbol{Hx}$. The difference between the two transformations is then presented, which is directly indicative of the linearization error arising from the Taylor series approximation. Figures~\ref{RcloseT}~and~\ref{RfarT} show the linearization errors of range measurements at close and far BS-UE ranges, respectively. It is clear that the linearization errors due to linearized transformation do increase in magnitude the closer the UE drives near the BS. Additional analysis on the errors can be found in \cite{DTCM}. Fig.~\ref{AT} depicts the transformation errors in angle when the linearization was done around $\theta=45\degree$. The figure shows a couple of findings. First, linearizing an arc-tan function through the Taylor series approximation would result in a considerable amount of errors that can reach up to $240\degree$. Second, it can be seen that the linearization error was at its lowest when $\Delta x$ and $\Delta y$ are equal, which is expected as the linearization was performed around $45\degree$. Third, linearization of the arc-tan function is very sensitive to any errors in the innovation sequence, as linearization errors would dramatically increase when the innovation sequence errors cease to be very close to zero. Hence, it is not advised to linearize such a function if other alternative solutions are available, which is the case with decentralized integration schemes.

\subsection{Proposed Decentralized Integration}
Decentralized integration of multiple BSs is viewed by the author as one of the main advantages of 5G-NR. Decentralized integration schemes were not attainable for GNSS-based systems in the past, as individual satellites do not have the ability to estimate the position of the UE using a single range measurement. Thus, GNSS systems were forced to use centralized fusion schemes along with their inevitable linearization errors. On the contrary, individual 5G BSs are capable of providing independent estimates of the position of the UE, enabling decentralized integration of BSs, where the relationship between the states and the measurements is linear. In order to compute the 3D position of a vehicle for decentralized fusion using a single BS, the range, the azimuth angle, and the elevation angles are required as seen below:

\begin{equation}
    \begin{split}
        &x=r\cdot \sin(\theta) \cos(\phi) + x_{BS}\\
        &y=r\cdot \cos(\theta) \cos(\phi) + y_{BS}\\
        &z=r\cdot \sin(\phi) +z_{BS},
    \end{split}
\end{equation}

\noindent where $\phi$ is the elevation angle. To measure the elevation angle along with the azimuth angle of departure, a 2D uniform rectangular array (URA) of antennas would be required. Alternatively, if the height of the vehicle is known to be constant, then the requirement for knowledge of elevation angle can be alleviated. Thus, requiring only a 1D uniform linear array (ULA). Such an assumption usually holds for land vehicles, as changes in the z-axis are generally minute. The position of the vehicle can thus be computed as follows:

\begin{figure}[t!]
	\centering
	\includegraphics[width=1\columnwidth]{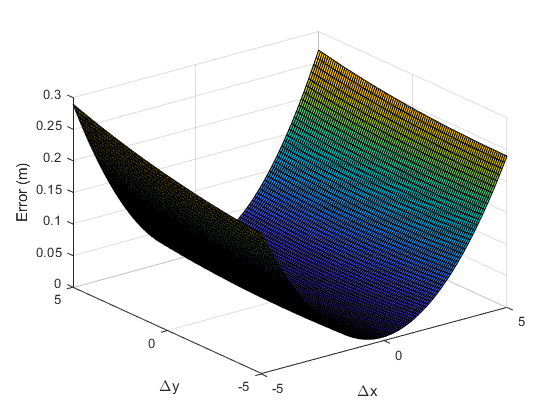}
	\DeclareGraphicsExtensions.
	\caption{Range linearization error at far BS-UE range ($\Delta x=1m, \Delta y=50$ m).}
	\label{RfarT}
\end{figure}

\begin{equation}\label{2DPos}
    \begin{split}
        &x=r_{2D} \cdot \sin(\theta) +x_{BS}\\
        &y=r_{2D} \cdot \cos(\theta) +y_{BS}\\
        &r_{2D}=\sqrt{r^2-\Delta z^2},
    \end{split}
\end{equation}

\noindent where $r_{2D}$ is the 2D range between the BS and the UE, and $\Delta z$ is the known height difference between the BS and the UE. It can be seen that the computations in (\ref{2DPos}) would still utilize non-linear models. Yet, such non-linear models are utilized outside of the KF implementation. Hence, no linearization is required.

In the proposed decentralized 5G multi-BS sensor fusion, the measurement vector $\boldsymbol{z}$ will be as follows: $\boldsymbol{z}~=~[x_1, y_1, \dots, x_N, y_N]^T$. It can be seen that the linear relationship between the states and measurements is indeed linear. Therefore, the $\boldsymbol{H}_k$ matrix for a decentralized fusion of 5G BSs, presented in (\ref{H}), can fully describe the relationship between the state vector $\boldsymbol{x}$ and the measurement vector $\boldsymbol{z}$, while not incurring any linearization errors.

\begin{equation} \label{H}
\boldsymbol{H}=\begin{bmatrix}
1 & 0 & 0 & 0\\
0 & 1 & 0 & 0\\
\vdots & \vdots & \vdots & \vdots\\
1 & 0 & 0 & 0\\
0 & 1 & 0 & 0
\end{bmatrix}
\end{equation}

It is worth noting that despite the simplistic and elegant design of the proposed method, the 5G positioning literature has failed to propose such a scheme. Thus, in this paper, we propose adopting decentralized integration schemes for 5G multi-BS positioning as it eliminates linearization errors without incurring any downsides in return.

\section{Experimental Setup and Results}

\subsection{Experimental Setup}\label{Exp Setup}
\begin{figure}[t!]
	\centering
	\includegraphics[width=\columnwidth]{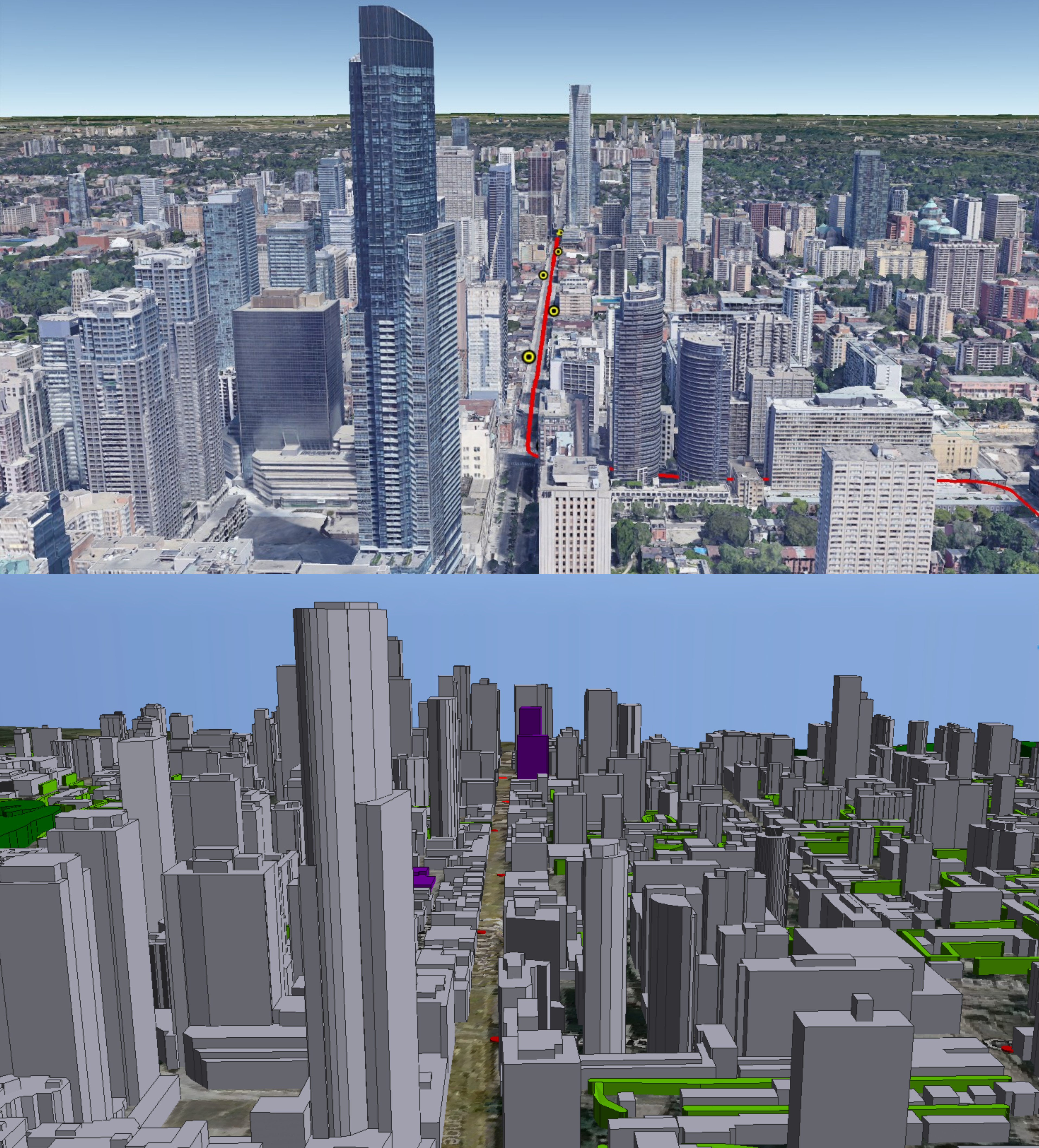}
	\DeclareGraphicsExtensions.
	\caption{Downtown Toronto, ON, Google Earth (Top) vs Siradel simulation tool (Bottom).}
	\label{GoogleEarth vs Siradel}
\end{figure}

\begin{figure}[t!]
	\centering
	\includegraphics[width=\columnwidth]{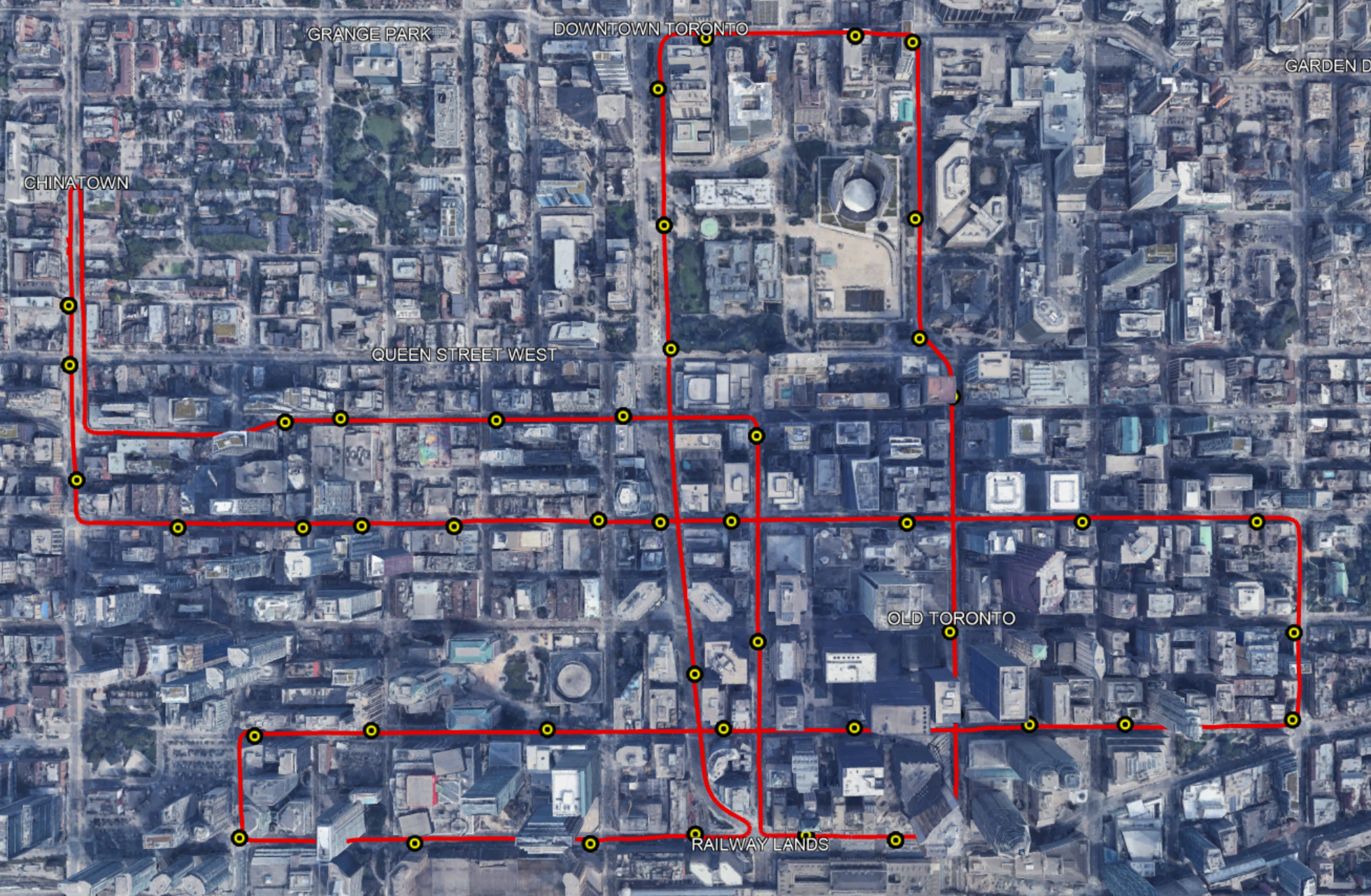}
	\DeclareGraphicsExtensions.
	\caption{Downtown Toronto Trajectory (Red), and 5G BSs (Yellow circles).}
	\label{Traj}
\end{figure}

In order to verify the superiority of decentralized/LC integration over centralized/TC integration, a quasi-real 5G simulation setup that was provided by Siradel was utilized. Siradel 5G\_Channel comprises LiDAR-based maps of the buildings, vegetation, and water bodies in downtown areas of cities like Toronto, New York, and Paris as seen in Fig.~\ref{GoogleEarth vs Siradel}. The simulation tool requires the position of the UE and the virtually connected BSs to compute required positioning measurables like RSS, TOA, AOA, and AOD via its ray-tracing capabilities and propagation models. To mimic a true urban navigation scenario, a vehicle was equipped with a high-end positioning solution provided by NovAtel; that includes KVH 1700, a tactical grade IMU, along with NovAtel's tactical grade GNSS receiver, and was driven in Downtown Toronto. The vehicle was driven for $1$ hr and $13$  mins, and the trajectory was $9$ km long as seen in Fig.~\ref{Traj}. The trajectory was conducted during rush hour, and thus, many sudden car stopping/acceleration dynamics were encountered. Moreover, the trajectory features many turns and challenging maneuvering dynamics. Next, a code was developed to place BSs along the driven trajectory that are $250$ m apart, as per 3GPP's Release 16 guidelines \cite{Course}. Finally, the generated BS positions and NovAtel's reference solution were imported to Siradel to generate the needed 5G measurables. Siradel was set up to utilize mmWave signals with a carrier frequency around $28$ GHz and a bandwidth of $400$ MHz. The BSs were equipped with $8\times1$ ULAs while the UE was equipped with an omnidirectional antenna.



\subsection{Results and Discussions}
In this section, the performance of the centralized schemes using EKF and UKF will be compared to the performance of the proposed decentralized scheme using a KF. The comparison will be done in three scenarios. First, perfect range and angle measurements are supplied to all filters. Positioning errors arising from this test will be solely attributed to linearization errors. The second test scenario will incorporate quasi-real measurements from Siradel's simulations and NLOS communications will be permitted without NLOS detection functionalities. Such a test will show the need for NLOS detection capabilities as a prerequisite to lane-level positioning. Finally, the third test scenario will include quasi-real measurements and NLOS detection capabilities discussed in \cite{NLOS}. This will act as an accurate benchmark of the expected performance of the compared filters and integration schemes. For all test scenarios, a trilateration variant solution will also be presented for EKF and UKF solutions; referred to as EKF-R and UKF-R, to gauge the advantages and disadvantages of adding angle measurements to centralized fusion.

\begin{figure}[t!]
	\centering
	\includegraphics[trim=117pt 241 125pt 250pt,clip,width=\columnwidth]{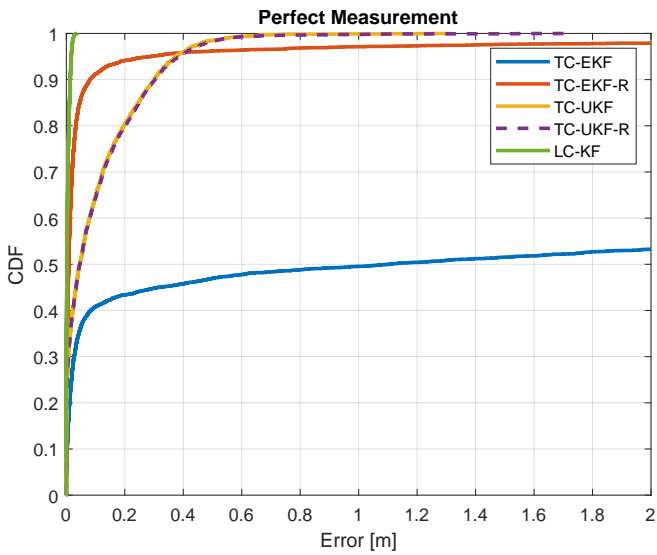}
	\DeclareGraphicsExtensions.
	\caption{CDF of the 2D positioning errors while utilizing perfect range and angle measurements.}
	\label{CDF-P}
\end{figure}

\begin{figure}[t!]
	\centering
	\includegraphics[width=\columnwidth]{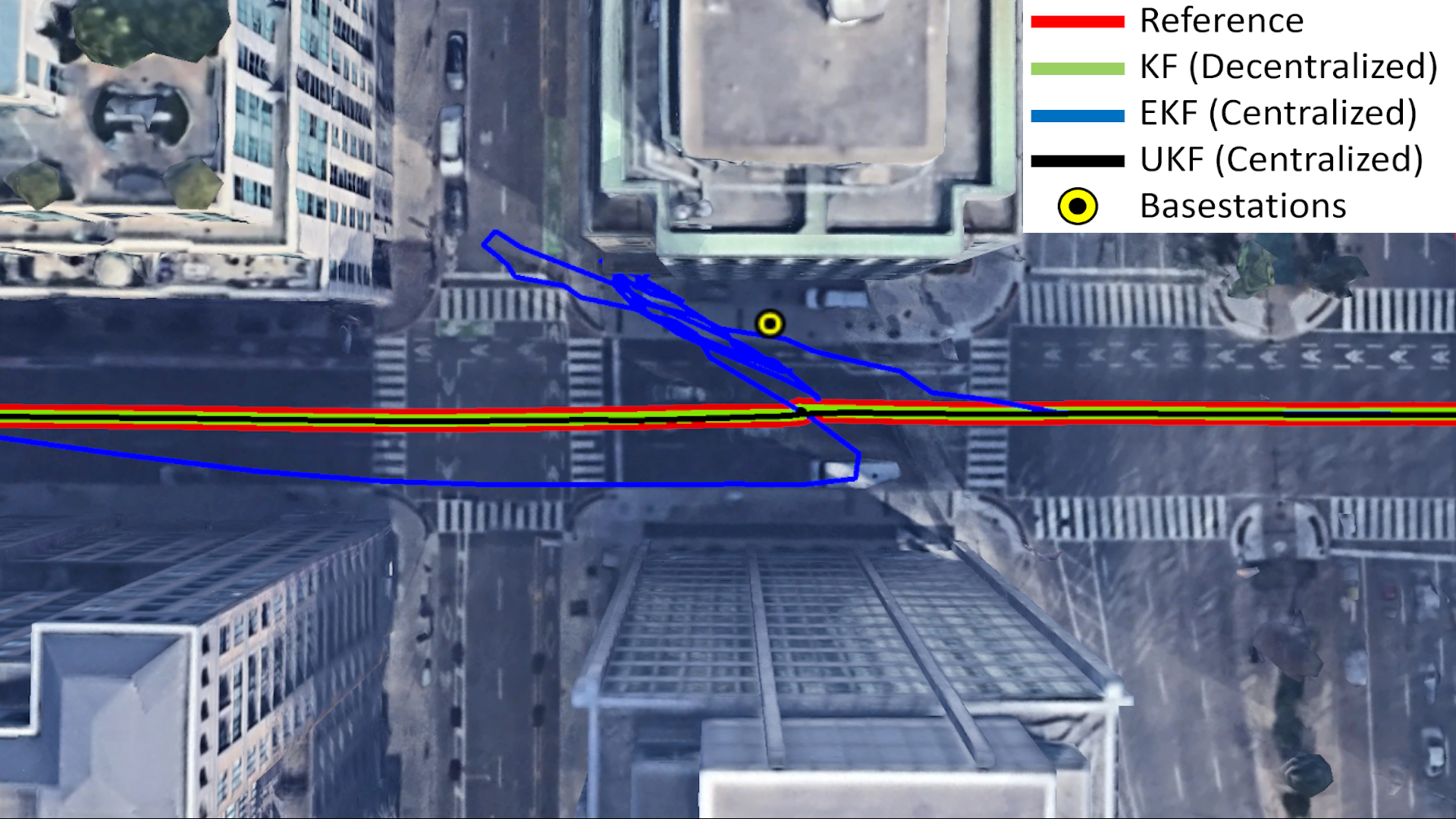}
	\DeclareGraphicsExtensions.
	\caption{First close-up to showcase the performance of filters while driving close to a BS using perfect measurement}
	\label{Perfect1}
\end{figure}

\begin{figure}[t!]
	\centering
	\includegraphics[width=\columnwidth]{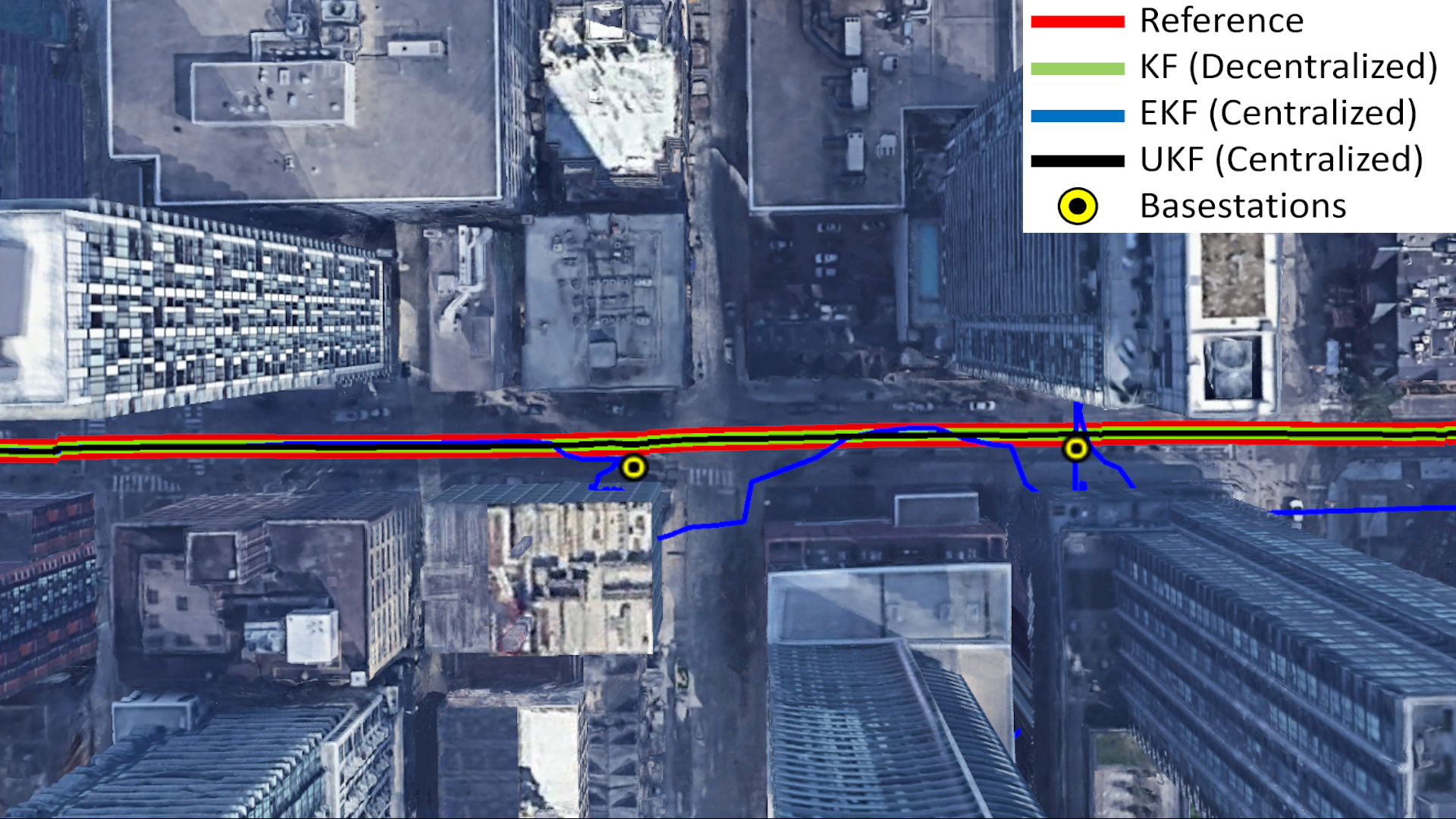}
	\DeclareGraphicsExtensions.
	\caption{Second close-up to showcase the performance of filters while driving close to a BS using perfect measurements.}
	\label{Perfect2}
\end{figure}

Fig.~\ref{CDF-P} depicts the Cumulative Distribution Function (CDF) of all implemented filters and integration schemes using perfect measurements. It is evident that the decentralized KF implementation trumped all other schemes, as it had an almost perfect performance. On the other hand, it can be seen that the EKF hybrid implementation had the lowest performance, even compared to its trilateration counterpart. This proves that linearizing angle-based measurements hinder the performance of the filter as it adds more errors than extra information. Additionally, both UKF implementations had almost similar performance, meaning that adding angle measurements did not add extra value in centralized fusion schemes. Moreover, it is apparent that the UKF had more errors at the decimeter level compared to its EKF counterpart, yet, it was able to suppress errors above $0.5$ m. To have a deeper insight into how linearization errors affect measurements, two close-up test samples are shown in Figures \ref{Perfect1} and \ref{Perfect2}, where the UE drives close to BSs. It can be seen that both decentralized KF and centralized UKF did not sustain much error while passing by nearby BSs. On the contrary, the centralized EKF implementation has sustained considerable positioning errors due to linearization errors.

\begin{figure}[t!]
	\centering
	\includegraphics[trim=117pt 241 125pt 245pt,clip,width=\columnwidth]{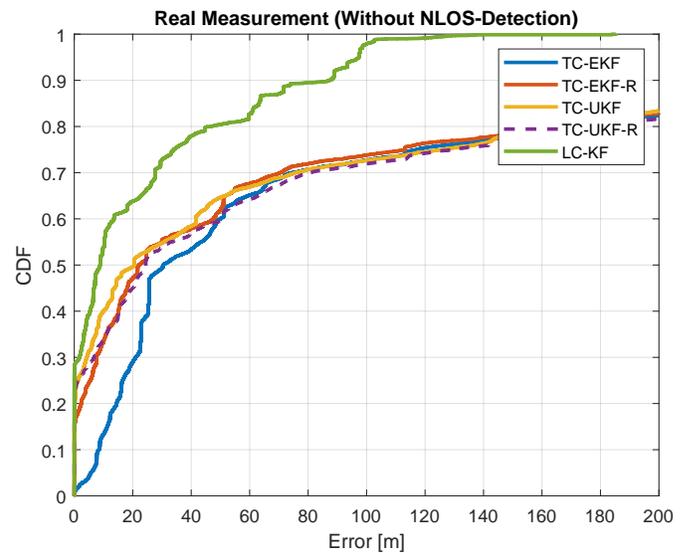}
	\DeclareGraphicsExtensions.
	\caption{CDF of the 2D positioning errors while utilizing quasi-real Siradel range and angle measurements without NLOS detection.}
	\label{CDF-R}
\end{figure}

Fig.~\ref{CDF-R} illustrates the CDF of all implemented filters and integration schemes using real range and angle measurements from Siradel without NLOS detection capabilities. Patently, the proposed decentralized KF implementation has outperformed all other schemes by a remarkable margin. Yet, such performance is not enough to achieve lane-level accuracy, as it was not able to suppress high levels of errors. Meanwhile, all other implementations failed to provide presentable estimations, as they sustained sub-2m level of accuracy less than 26\% of the time. This is expected, as linearization errors have a higher impact when the innovation sequence's residual error is high in value, which is the case for NLOS measurements. This finding shows the importance of the development of NLOS detection algorithms, which are not given enough attention in the community. Also, it proves the superiority of the proposed decentralized approach in case the NLOS detector failed to capture an NLOS communication link.

\begin{figure}[t!]
	\centering
	\includegraphics[trim=117pt 241 125pt 245pt,clip,width=\columnwidth]{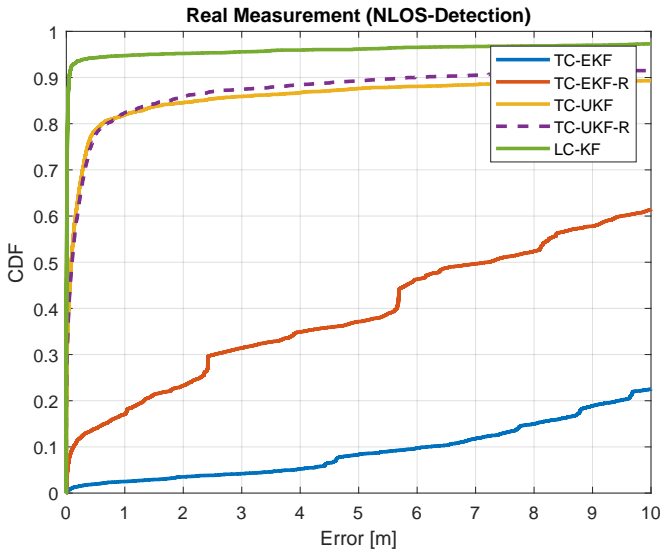}
	\DeclareGraphicsExtensions.
	\caption{CDF of the 2D positioning errors while utilizing quasi-real Siradel range and angle measurements with NLOS detection.}
	\label{CDF-NLOS}
\end{figure}

Fig.~\ref{CDF-NLOS} shows the CDF of the positioning error for all aforementioned schemes while using real Siradel range and angle measurements along with NLOS detection capabilities found in \cite{NLOS}. Evidently, the proposed decentralized KF implementation was able to outperform other implementations in both decimeter-level accuracy and meter-level accuracy statistics. The proposed method was able to navigate with a sub-meter level of accuracy for around $91\%$ of the time. Both UKF implementations were able to maintain a sub-1m level of accuracy for around $82\%$ of the time. It is also noticeable that the trilateration-based UKF was slightly better than its hybrid counterpart due to the avoidance of extra linearization errors imposed by the angle measurements. Finally, it is conspicuous beyond doubt that both EKF implementations are unsuitable for lane-level positioning, albeit the trilateration version had remarkably better performance for the aforementioned reasons.

\begin{figure}[t!]
	\centering
	\includegraphics[width=\columnwidth]{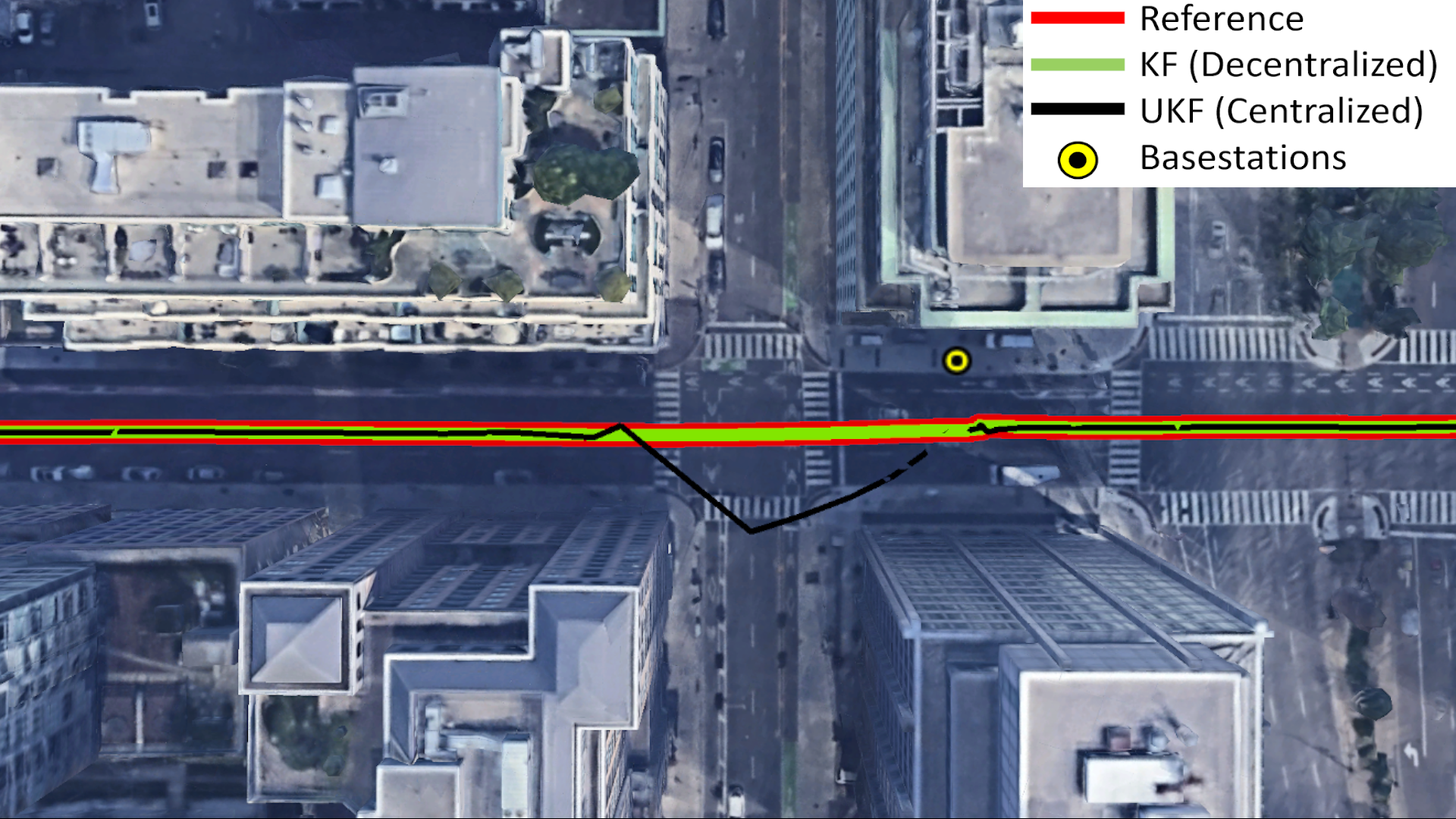}
	\DeclareGraphicsExtensions.
	\caption{First close-up to showcase the performance of filters while driving close to a BS while connected to multiple BSs and using quasi-real measurements.}
	\label{Real1}
\end{figure}

\begin{figure}[t!]
	\centering
	\includegraphics[width=\columnwidth]{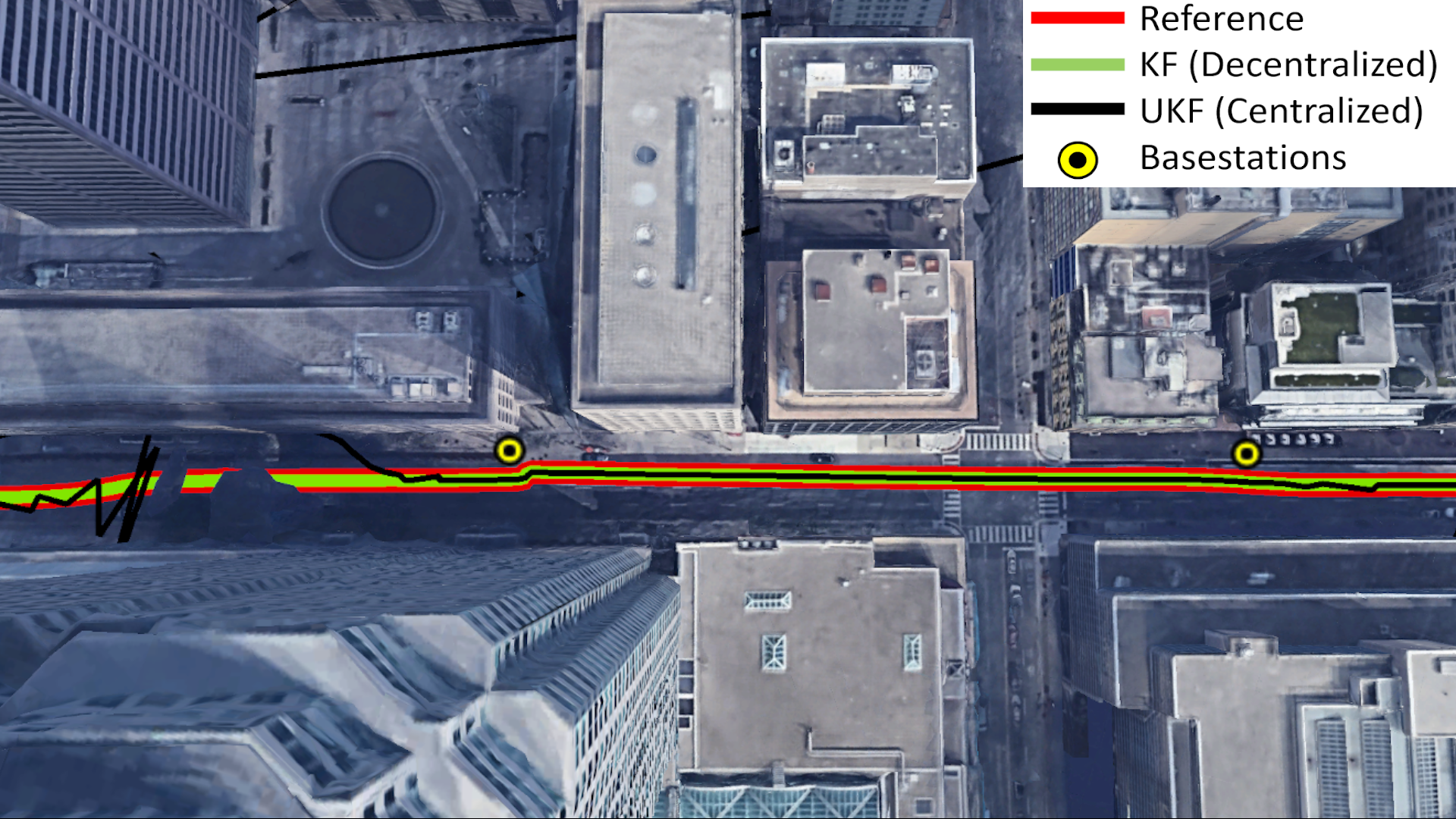}
	\DeclareGraphicsExtensions.
	\caption{Second close-up to showcase the performance of filters while driving close to a BS while connected to a single BS and using quasi-real measurements.}
	\label{Real2}
\end{figure}

\begin{figure}[t!]
	\centering
	\includegraphics[width=\columnwidth]{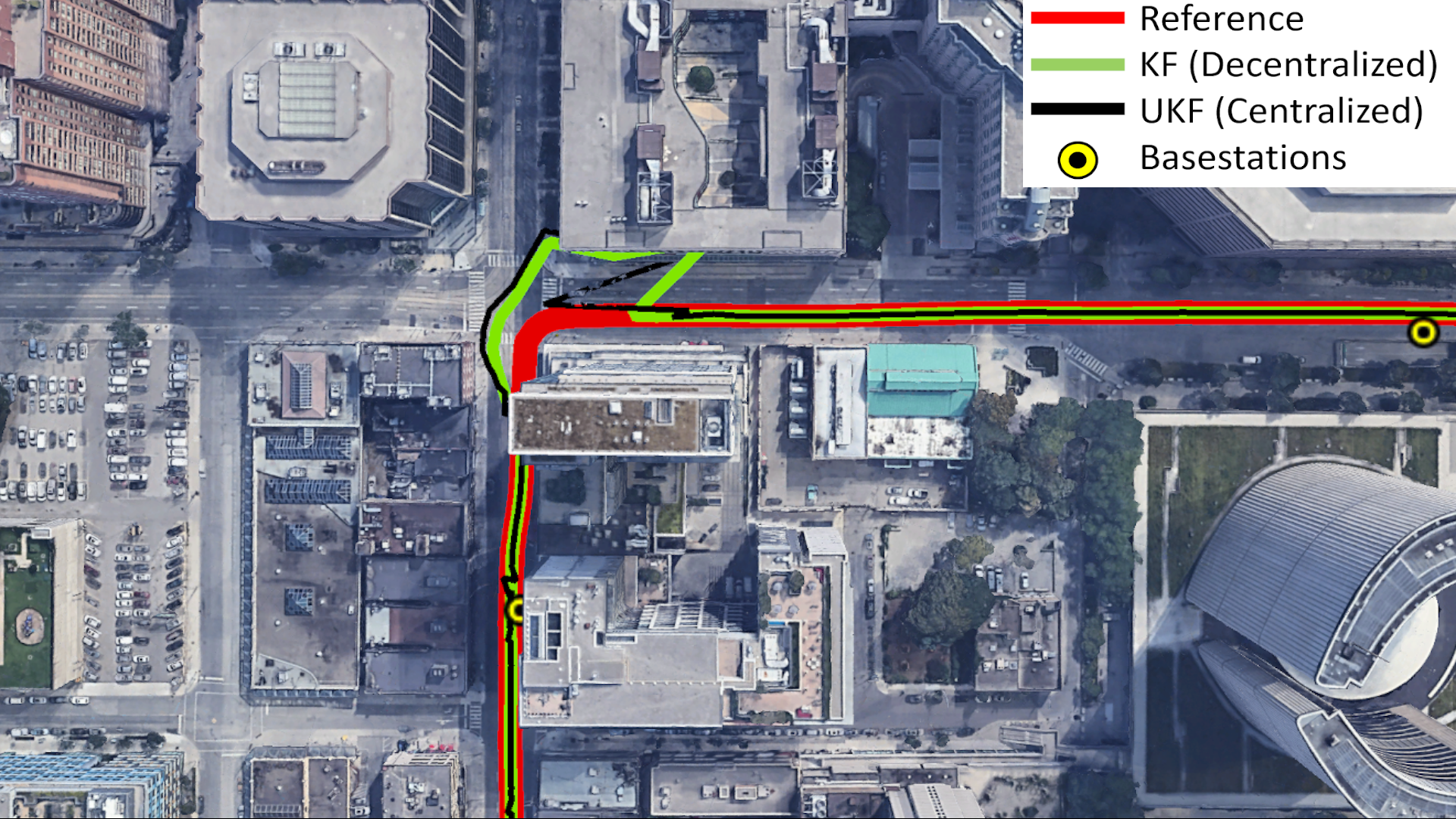}
	\DeclareGraphicsExtensions.
	\caption{Third close-up to showcase the performance of filters while driving close to a BS while connected to NLOS BSs and using quasi-real measurements.}
	\label{Real3}
\end{figure}

Figures \ref{Real1}-\ref{Real3} show close-up samples of the positioning solution of both decentralized KF and centralized UKF in three different scenarios. The first scenario, shown in Fig. \ref{Real1}, depicts the performance of the filters when close to a BS while connected to multiple BSs at the same time. It can be seen that the proposed decentralized KF method did not sustain errors while the centralized UKF implementation sustained lane-level errors for a short time around the close BS due to linearization errors. The second scenario, shown in Fig. \ref{Real2}, depicts the performance of the filters when they are connected to a single BS. The superiority of the proposed method over centralized UKF can be clearly seen, as the UKF implementation suffered from positioning errors for a longer period of time, due to the lack of supporting BSs. Finally, the third scenario, shown in Fig. \ref{Real3}, illustrates the performance when the UE is in total NLOS communication with all nearby BSs. It can be seen that both implementations have sustained large amounts of errors. Such a result is expected to occur from time to time while using 5G standalone positioning. Thus, the authors see a crucial need for integration between 5G and other onboard motion sensors to bridge such gaps. Investigation of novel integration methodologies between 5G and other sensors is to be conducted in future works. A summary of the performance statistics of all algorithms and schemes is found in Table \ref{Real_resultswith}.

\begin{table}[ht]
	\caption{2D Positioning Error Statistics Summary}
	\label{Real_resultswith}
	\begin{tabularx}{\columnwidth}{@{}l*{5}{C}c@{}}
		\toprule
		Statistics &LC-KF 	        &TC-UKF\hspace{2pt}        &\hspace{-6pt}TC-UKF-R      &TC-EKF        &\hspace{-5pt}TC-EKF-R\\
		\midrule
		RMS        & 9 m       &47.7 m    &12.8 m     &881.2 m   &115.7 m\\ 
		Max        & 89.3 m    &885.7 m   &145.3 m    &5.8e4 m   &2.3e3 m\\
		$<2$ m     & 95.2\%    &84.5\%    &85.7\%     &3.5\%     &23.3\%\\ 
		$<1$ m     & 94.7\%    &81.9\%    &82.4\%     &2.5\%     &17.1\%\\
		$<30$ cm   & 93.9\%    &71.4\%    &68.9\%     &1.7\%     &12.7\%\\
		\bottomrule
	\end{tabularx}
\end{table}

\section{Conclusion}
In conclusion, it was shown that 5G multi-BS integration schemes for positioning purposes in the literature are heavily inspired by GNSS trilateration centralized schemes. It was proven that such schemes suffer from linearization errors due to the direct use of range and angle measurements. Such errors were necessary for GNSS-based algorithms, as the states and the measurements do not have a one-to-one function, as they rely on range measurements only. In 5G, however, it was shown that with the aid of angle-based measurements, a one-to-one function between the states and the measurements can be achieved, opening the way for integration on the position level rather than on the measurement level. Therefore, it was proposed to adopt a decentralized KF-based integration scheme,  which totally eliminates linearization errors. To test the proposed integration scheme, a quasi-real simulation system was utilized to simulate 5G operation in downtown Toronto for a real trajectory that is $1$ hour and $13$ minutes long. The tests took place in a perfect measurement scenario and in a real measurement scenario with/without NLOS detection capabilities. It was shown that the decentralized positioning scheme was able to outperform its centralized counterpart schemes that utilize EKF-based and UKF-based filters by a considerable margin in all scenarios.

\bibliographystyle{IEEEtran}
\bibliography{references,references_Zotero}

\begin{thebibliography}{10}
\providecommand{\url}[1]{#1}
\csname url@samestyle\endcsname
\providecommand{\newblock}{\relax}
\providecommand{\bibinfo}[2]{#2}
\providecommand{\BIBentrySTDinterwordspacing}{\spaceskip=0pt\relax}
\providecommand{\BIBentryALTinterwordstretchfactor}{4}
\providecommand{\BIBentryALTinterwordspacing}{\spaceskip=\fontdimen2\font plus
\BIBentryALTinterwordstretchfactor\fontdimen3\font minus
  \fontdimen4\font\relax}
\providecommand{\BIBforeignlanguage}[2]{{%
\expandafter\ifx\csname l@#1\endcsname\relax
\typeout{** WARNING: IEEEtran.bst: No hyphenation pattern has been}%
\typeout{** loaded for the language `#1'. Using the pattern for}%
\typeout{** the default language instead.}%
\else
\language=\csname l@#1\endcsname
\fi
#2}}
\providecommand{\BIBdecl}{\relax}
\BIBdecl

\bibitem{Reid_2019}
\BIBentryALTinterwordspacing
T.~G. Reid, S.~E. Houts, R.~Cammarata, G.~Mills, S.~Agarwal, A.~Vora, and
  G.~Pandey, ``Localization requirements for autonomous vehicles,'' \emph{{SAE}
  International Journal of Connected and Automated Vehicles}, vol.~2, no.~3,
  sep 2019. [Online]. Available: \url{https://doi.org/10.4271\%2F12-02-03-0012}
\BIBentrySTDinterwordspacing

\bibitem{5GPos}
F.~Mogyorósi, P.~Revisnyei, A.~Pašić, Z.~Papp, I.~Törös, P.~Varga, and
  A.~Pašić, ``Positioning in {5G} and {6G} networks---a survey,''
  \emph{Sensors}, vol.~22, no.~13, 2022.

\bibitem{RTK}
P.~Nair and A.~S. Sarif, ``A review on application of global positioning system
  in construction industry,'' \emph{Progress in Engineering Application and
  Technology}, vol.~3, no.~1, p. 249–259, Jun. 2022.

\bibitem{UWB}
F.~Mazhar, M.~G. Khan, and B.~S\"allberg, ``Precise indoor positioning using
  {UWB}: A review of methods, algorithms and implementations,'' \emph{Wireless
  Personal Communications}, vol.~97, no.~3, p. 4467–4491, 2017.

\bibitem{WiFi}
F.~Liu, J.~Liu, Y.~Yin, W.~Wang, D.~Hu, P.~Chen, and Q.~Niu, ``Survey on
  {WiFi}-based indoor positioning techniques,'' \emph{IET Communications},
  vol.~14, no.~9, pp. 1372--1383, 2020.

\bibitem{merits}
S.~Dwivedi, R.~Shreevastav, F.~Munier, J.~Nygren, I.~Siomina, Y.~Lyazidi,
  D.~Shrestha, G.~Lindmark, P.~Ernstr{\"o}m, E.~Stare, S.~M. Razavi,
  S.~Muruganathan, G.~Masini, {\AA}.~Busin, and F.~Gunnarsson, ``Positioning in
  {5G} networks,'' \emph{IEEE Communications Magazine}, vol.~59, no.~11, pp.
  38--44, 2021.

\bibitem{KF}
R.~E. K{\'a}lm{\'a}n, ``A new approach to linear filtering and prediction
  problems,'' \emph{Journal of Basic Engineering}, vol.~82, no.~1, pp. 35--45,
  03 1960.

\bibitem{EKF}
H.~W. Sorenson and A.~R. Stubberud, ``Non-linear filtering by approximation of
  the a posteriori density,'' \emph{International Journal of Control}, vol.~8,
  no.~1, p. 33–51, 1968.

\bibitem{ukf}
S.~Julier and J.~Uhlmann, ``Unscented filtering and nonlinear estimation,''
  \emph{Proceedings of the IEEE}, vol.~92, no.~3, pp. 401--422, 2004.

\bibitem{TrainTOAOA}
J.~Talvitie, T.~Levanen, M.~Koivisto, K.~Pajukoski, M.~Renfors, and M.~Valkama,
  ``Positioning of high-speed trains using {5G} new radio synchronization
  signals,'' in \emph{IEEE Wireless Communications and Networking Conference
  (WCNC)}, 2018, pp. 1--6.

\bibitem{TrainTDOA1}
J.~Talvitie, T.~Levanen, M.~Koivisto, T.~Ihalainen, K.~Pajukoski, M.~Renfors,
  and M.~Valkama, ``Positioning and location-based beamforming for high speed
  trains in {5G} {NR} networks,'' in \emph{IEEE Globecom Workshops (GC
  Wkshps)}, 2018, pp. 1--7.

\bibitem{TrainTDOA2}
J.~Talvitie, T.~Levanen, M.~Koivisto, T.~Ihalainen, K.~Pajukoski, and
  M.~Valkama, ``Positioning and location-aware communications for modern
  railways with {5G} new radio,'' \emph{IEEE Communications Magazine}, vol.~57,
  no.~9, pp. 24--30, 2019.

\bibitem{TrainTDOAAOA}
J.~Talvitie, M.~Koivisto, T.~Levanen, T.~Ihalainen, K.~Pajukoski, and
  M.~Valkama, ``Radio positioning and tracking of high-speed devices in {5G}
  {NR} networks: System concept and performance,'' in \emph{European Signal
  Processing Conference (EUSIPCO)}, 2019, pp. 1--5.

\bibitem{TrainNonLinear}
J.~Talvitie, T.~Levanen, M.~Koivisto, and M.~Valkama, ``Positioning and
  tracking of high-speed trains with non-linear state model for {5G} and beyond
  systems,'' in \emph{International Symposium on Wireless Communication Systems
  (ISWCS)}, 2019, pp. 309--314.

\bibitem{trivedi_localization_2021}
M.~A. Trivedi and J.~H. van Wyk, ``Localization and {Tracking} of {High}-speed
  {Trains} {Using} {Compressed} {Sensing} {Based} {5G} {Localization}
  {Algorithms},'' in \emph{{IEEE} {International} {Conference} on {Information}
  {Fusion} ({FUSION})}, Nov. 2021, pp. 1--8.

\bibitem{5GHybrid2}
N.~{Garcia}, H.~{Wymeersch}, E.~G. {Larsson}, A.~M. {Haimovich}, and
  M.~{Coulon}, ``Direct localization for massive {MIMO},'' \emph{IEEE
  Transactions on Signal Processing}, vol.~65, no.~10, pp. 2475--2487, 2017.

\bibitem{rastorgueva-foi_beam-based_2018}
\BIBentryALTinterwordspacing
E.~Rastorgueva-Foi, M.~Costa, M.~Koivisto, J.~Talvitie, K.~Leppaneny, and
  M.~Valkama, ``\BIBforeignlanguage{en}{Beam-based {Device} {Positioning} in
  {mmWave} {5G} {Systems} under {Orientation} {Uncertainties}},'' in
  \emph{\BIBforeignlanguage{en}{Asilomar {Conference} on {Signals}, {Systems},
  and {Computers}}}.\hskip 1em plus 0.5em minus 0.4em\relax Pacific Grove, CA,
  USA: IEEE, Oct. 2018, pp. 3--7. [Online]. Available:
  \url{https://ieeexplore.ieee.org/document/8645340/}
\BIBentrySTDinterwordspacing

\bibitem{gertzell_5g_2020}
P.~Gertzell, J.~Landelius, H.~Nyqvist, A.~Fascista, A.~Coluccia,
  G.~Seco-Granados, N.~Garcia, and H.~Wymeersch, ``{5G} multi-{BS}
  {Positioning} with a {Single}-{Antenna} {Receiver},'' in \emph{{IEEE}
  {Annual} {International} {Symposium} on {Personal}, {Indoor} and {Mobile}
  {Radio} {Communications} ({PIMRC})}, Aug. 2020, pp. 1--5, iSSN: 2166-9589.

\bibitem{ko_high-speed_2022}
K.~Ko, I.~Byun, W.~Ahn, and W.~Shin, ``High-{Speed} {Train} {Positioning}
  {Using} {Deep} {Kalman} {Filter} {With} {5G} {NR} {Signals},'' \emph{IEEE
  Transactions on Intelligent Transportation Systems}, vol.~23, no.~9, pp.
  15\,993--16\,004, Sep. 2022, conference Name: IEEE Transactions on
  Intelligent Transportation Systems.

\bibitem{xhafa_evaluation_2022}
\BIBentryALTinterwordspacing
A.~Xhafa, J.~A. del Peral-Rosado, J.~A. López-Salcedo, and G.~Seco-Granados,
  ``\BIBforeignlanguage{en}{Evaluation of {5G} {Positioning} {Performance}
  {Based} on {UTDoA}, {AoA} and {Base}-{Station} {Selective} {Exclusion}},''
  \emph{\BIBforeignlanguage{en}{Sensors}}, vol.~22, no.~1, p. 101, Jan. 2022,
  number: 1 Publisher: Multidisciplinary Digital Publishing Institute.
  [Online]. Available: \url{https://www.mdpi.com/1424-8220/22/1/101}
\BIBentrySTDinterwordspacing

\bibitem{5GSync2}
M.~Koivisto, M.~Costa, A.~Hakkarainen, K.~Leppanen, and M.~Valkama, ``Joint
  {3D} positioning and network synchronization in {5G} ultra-dense networks
  using {UKF} and {EKF},'' in \emph{IEEE Globecom Workshops (GC Wkshps)}, 2016,
  pp. 1--7.

\bibitem{whitepaper}
{5G~Americas}, ``{Cellular Communication In a {5G} Era},'' 5G Americas, Tech.
  Rep., 01 2022.

\bibitem{5GSync1}
J.~Werner, M.~Costa, A.~Hakkarainen, K.~Leppanen, and M.~Valkama, ``Joint user
  node positioning and clock offset estimation in {5G} ultra-dense networks,''
  in \emph{IEEE Global Communications Conference (GLOBECOM)}, 2015, pp. 1--7.

\bibitem{bai_toa-aoa_2021}
L.~Bai, C.~Sun, H.~Zhao, J.~W. Cheong, A.~G. Dempster, and W.~Feng,
  ``\BIBforeignlanguage{en}{A {TOA}-{AOA} {Hybrid} {Localization} {Method} in
  {5G} {Network} with {MIMO} {Antennas}},'' in
  \emph{\BIBforeignlanguage{en}{China {Satellite} {Navigation} {Conference}
  ({CSNC})}}, ser. Lecture {Notes} in {Electrical} {Engineering}, C.~Yang and
  J.~Xie, Eds.\hskip 1em plus 0.5em minus 0.4em\relax Singapore: Springer,
  2021, pp. 285--295.

\bibitem{rastorgueva-foi_localization_2018}
\BIBentryALTinterwordspacing
E.~Rastorgueva-Foi, M.~Koivisto, M.~Valkama, M.~Costa, and K.~Leppanen,
  ``\BIBforeignlanguage{en}{Localization and {Tracking} in {mmWave} {Radio}
  {Networks} using {Beam}-{Based} {DoD} {Measurements}},'' in
  \emph{\BIBforeignlanguage{en}{International {Conference} on {Localization}
  and {GNSS} ({ICL}-{GNSS})}}.\hskip 1em plus 0.5em minus 0.4em\relax
  Guimaraes: IEEE, Jun. 2018, pp. 1--6. [Online]. Available:
  \url{https://ieeexplore.ieee.org/document/8440914/}
\BIBentrySTDinterwordspacing

\bibitem{shahmansoori_tracking_2019}
A.~Shahmansoori, B.~Uguen, G.~Destino, G.~Seco-Granados, and H.~Wymeersch,
  ``Tracking {Position} and {Orientation} {Through} {Millimeter} {Wave} {Lens}
  {MIMO} in {5G} {Systems},'' \emph{IEEE Signal Processing Letters}, vol.~26,
  no.~8, pp. 1222--1226, Aug. 2019, conference Name: IEEE Signal Processing
  Letters.

\bibitem{5GSync3}
M.~Koivisto, M.~Costa, J.~Werner, K.~Heiska, J.~Talvitie, K.~Lepp\"anen,
  V.~Koivunen, and M.~Valkama, ``Joint device positioning and clock
  synchronization in {5G} ultra-dense networks,'' \emph{IEEE Transactions on
  Wireless Communications}, vol.~16, no.~5, pp. 2866--2881, 2017.

\bibitem{Course}
J.~{Levine}, ``Fundamentals of {5G} small cell deployments,'' in \emph{IEEE
  Communications Society Training Course}, 2020.

\bibitem{NLOS}
Q.~Bader, S.~Saleh, M.~Elhabiby, and A.~Noureldin, ``{NLoS} detection for
  enhanced {5G mmWave}-based positioning for vehicular {IoT} applications,'' in
  \emph{2022 IEEE Globecom}, 2022.

\bibitem{DTCM}
S.~Saleh, S.~Sorour, and A.~Noureldin, ``Vehicular positioning using {mmWave}
  {TDOA} with a dynamically tuned covariance matrix,'' in \emph{IEEE Globecom
  Workshops (GC Wkshps)}, 2021, pp. 1--6.

\end{thebibliography}

\begin{IEEEbiography}[{\includegraphics[width=1in,height=1.25in,clip]{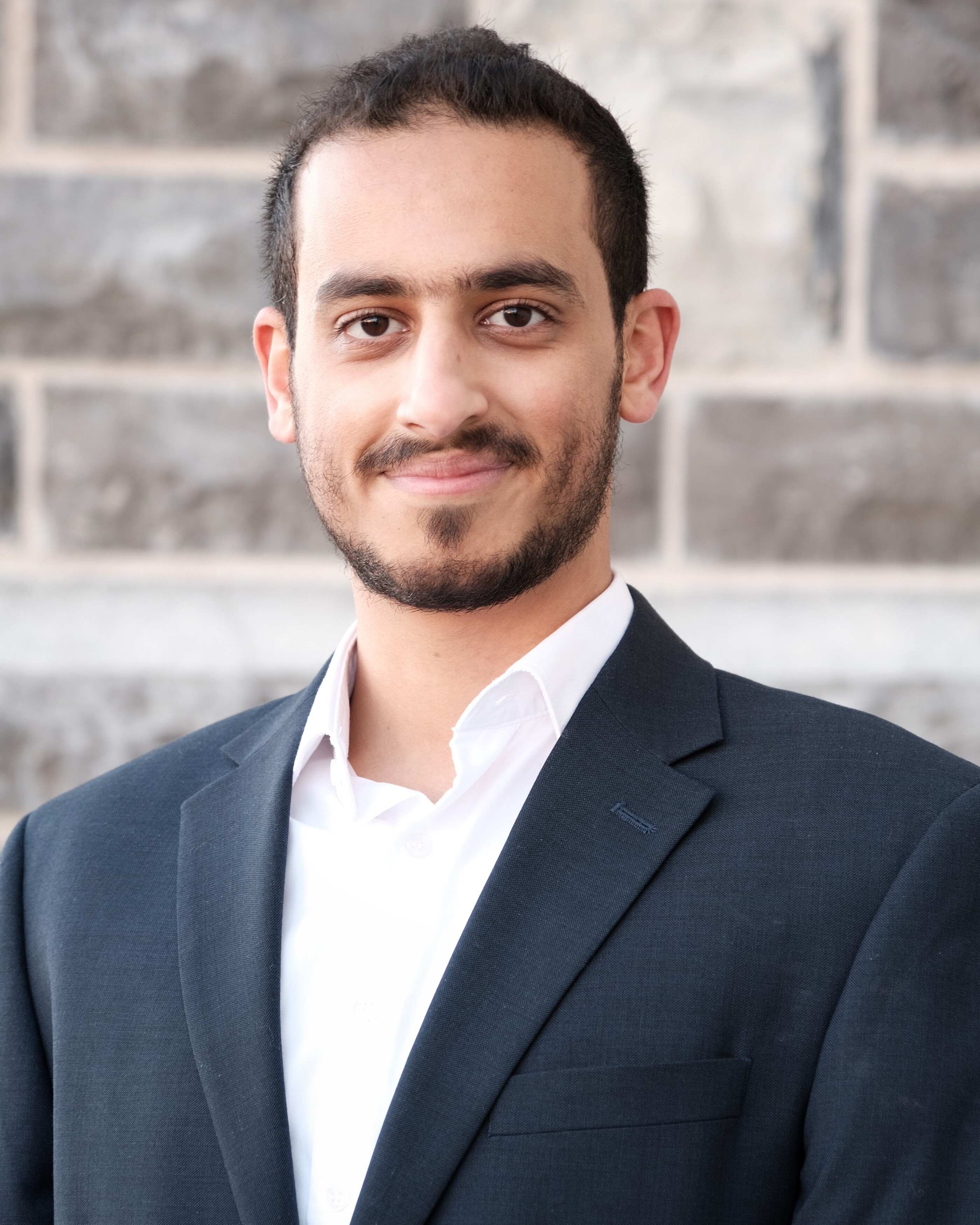}}]{Sharief Saleh} (Graduate Student Member, IEEE) received the B.Sc. and M.Sc. degrees in electrical engineering from Qatar University, Doha, Qatar, in 2016 and 2018, respectively, and is currently pursuing a PhD degree in electrical engineering at Queen's University, Canada. He was awarded a Graduate Assistant position at Qatar University during his master's studies and was then appointed as a Research Assistant at Qatar University, Doha, Qatar. He is currently a member of the Navigation and Instrumentation (NavINST) Research Lab, RMCC. His current research interests include 5G positioning and navigation, sensor fusion, sensors and instrumentation, estimation theory, and reinforcement learning.
\end{IEEEbiography}

\begin{IEEEbiography}[{\includegraphics[width=1in,height=1.25in,clip]{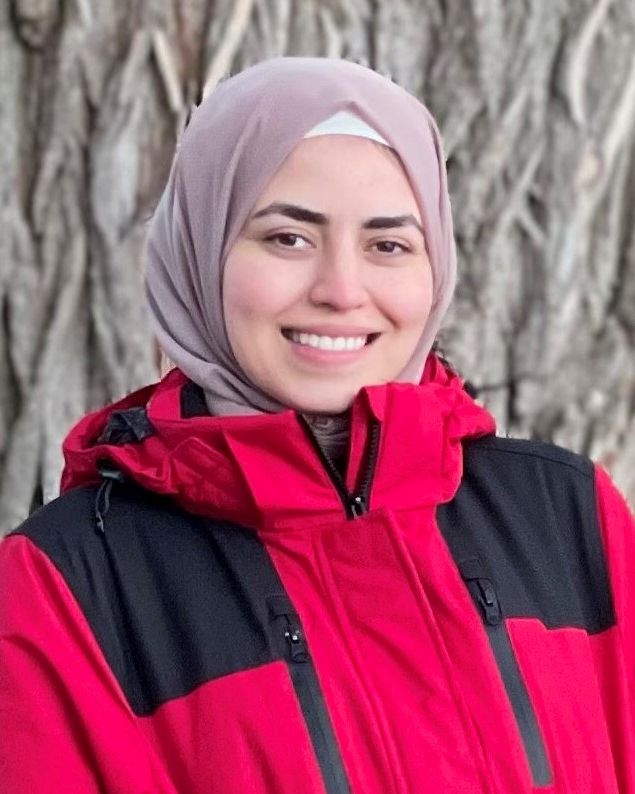}}]{Qamar Bader} (Graduate Student Member, IEEE) received her B.Sc. degree in electrical engineering from Qatar University, Doha, Qatar, in 2016 and is currently pursuing an MSc degree in electrical engineering at Queen's University, Canada. She is currently a member of the Navigation and Instrumentation (NavINST) Research Lab, RMCC. Her current research interests include 5G positioning and navigation, sensor fusion, deep learning, computer vision, and environment mapping.
\end{IEEEbiography}

\begin{IEEEbiography}[{\includegraphics[width=1in,height=1.25in,clip]{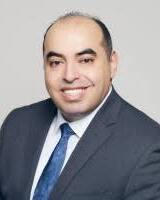}}]{Mohamed Elhabiby} was the Treasurer of the Geodesy Section at the Canadian Geophysical Union from 2008 to 2014. He is currently an Associate Professor with the Faculty of Engineering, Ain Shams University, Cairo, Egypt. He is also the Executive Vice President and a Co-Founder of Micro Engineering Tech Inc., Calgary, AB, a high-tech international company, specialized in high precision engineering and instrumentation, mobile mapping, laser scanning, deformation monitoring, and GPS/INS integrations. He is a Leader of an Archaeological Mission in the Area of Great Pyramids, Cairo. He received the Astech Awards. He is named by Avenue Magazine as one of the Top 40 under 40. He is the Chair of WG 4.1.4: Imaging Techniques, Sub-Commission 4.1: Alternatives and Backups to GNSS. He chaired the Geocomputations and Cyber Infrastructure Oral Session at the Canadian Geophysical Union annual meeting from 2008 to 2012.
\end{IEEEbiography}

\begin{IEEEbiography}[{\includegraphics[width=1in,height=1.25in,clip]{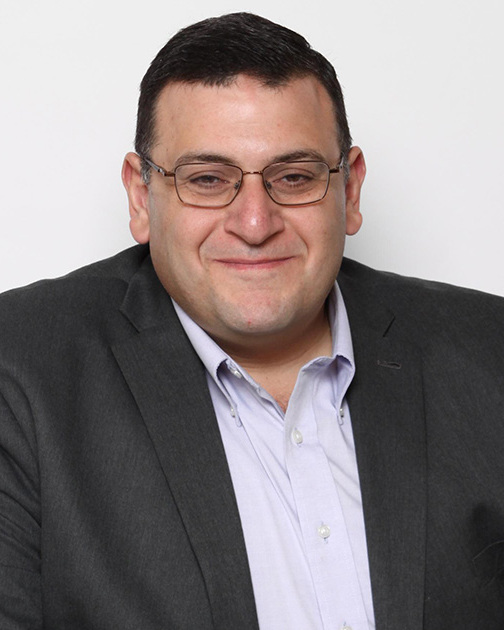}}]{Aboelmagd Noureldin} (Senior Member, IEEE) received the B.Sc. degree in electrical engineering and the M.Sc. degree in engineering physics from Cairo University, Egypt, in 1993 and 1997, respectively, and the Ph.D. degree in electrical and computer engineering from the University of Calgary, AB, Canada, in 2002. He is currently a Professor with the Department of Electrical and Computer Engineering, Royal Military College of Canada (RMCC) with Cross Appointment at the School of Computing and the Department of Electrical and Computer Engineering, Queen’s University. He is also the Founder and the Director of the Navigation and Instrumentation (NavINST) Research Lab, RMCC. He has published two books, four book chapters, and more than 270 papers in journals, magazines, and conference proceedings. His research interests include global navigation satellite systems, wireless location, and navigation, indoor positioning, and multi-sensor fusion targeting applications related to autonomous systems, intelligent transportation, road information services, crowd management, and the vehicular Internet of Things. His research work led to 13 patents and several technologies licensed to the industry in the area of position, location, and navigation systems.
\end{IEEEbiography}
\vfill
\end{document}